\documentclass[%
reprint,
 amsmath,amssymb,
 aps,
longbibliography]{revtex4-2}

\usepackage{graphicx}
\usepackage{subfigure}
\usepackage{dcolumn}
\usepackage{bm}
\usepackage{epstopdf}
\usepackage{xcolor}
\usepackage{hyperref}
\usepackage{appendix} 
\usepackage{mathtools}
\usepackage{tablefootnote}
\usepackage{comment}
\usepackage{textcase}
\usepackage[english]{babel}

\begin{document}

\preprint{APS/123-QED}

\title{A Comprehensive Study of the Three-- and Four--Neutron Systems at Low Energies}

\author{Michael D. Higgins}
\email{higgin45@purdue.edu}
\affiliation{Department of Physics and Astronomy, Purdue University, West Lafayette, Indiana 47907 USA}
\author{Chris H. Greene}
\email{chgreene@purdue.edu}
\affiliation{Department of Physics and Astronomy, Purdue University, West Lafayette, Indiana 47907 USA}
\affiliation{Purdue Quantum Science and Engineering Institute, Purdue University, West Lafayette, Indiana 47907 USA}
\author{A. Kievsky}%
 \email{alejandro.kievsky@pi.infn.it}
 \author{M. Viviani}
  \email{michele.viviani@pi.infn.it}
\affiliation{%
 Instituto Nazionale di Fisica Nucleare, Largo Pontecorvo 3, 56127 Pisa, Italy
}%


\date{\today}

\begin{abstract}
This work presents further analysis of the three-- and four--neutron systems in the low energy regime using adiabatic hyperspherical methods. In our previous Phys. Rev. Lett. article \cite{PhysRevLett.125.052501}, the low--energy behavior of these neutron systems was treated in the adiabatic approximation, neglecting the off--diagonal non--adiabatic couplings. A thorough analysis of the density of states through a multi--channel treatment of the three--and four--neutron scattering near the scattering continuum threshold is performed, showing no evidence of a 4n resonance at low energy. A detailed analysis of the long--range behavior of the lowest few adiabatic hyperspherical potentials shows there is an attractive $\rho^{-3}$ universal behavior which dominates in the low--energy regime of the multi--channel scattering. This long--range behavior leads to a divergent behavior of the density of state for $E\rightarrow0$ that could account for the low--energy signal observed in the 2016 experiment by Kisamori et al. \cite{Kisamori}.

\end{abstract}

\maketitle


\section{Introduction}

Few neutron systems have been the subject of interest over the past couple decades due to the possibility of forming low-energy bound states or temporary bound states through a long--lived resonance in the scattering continuum. The interest in these systems arose from experimental evidence of a possible low--energy tetraneutron (4n) state, most notably from experiments performed by Marqués et al. in 2002 \cite{PhysRevC.65.044006} in the reaction $\prescript{14}{4}{\mathrm{Be}}\rightarrow\prescript{10}{4}{\mathrm{Be}}+\prescript{}{}{\mathrm{4n}}$ and in a more recent experiment by Kisamori et al. in 2016 \cite{Kisamori} in the nuclear reaction $\prescript{4}{2}{\mathrm{He}}+\prescript{8}{2}{\mathrm{He}}\rightarrow\prescript{8}{4}{\mathrm{Be}}+\prescript{}{}{\mathrm{4n}}$. The interpretation of the results in these experiments sparked numerous theoretical studies into whether four interacting neutrons can bind or produce a resonance with current nuclear models \cite{Timofeyuk_2003,PhysRevLett.90.252501,Bertulani_2003,ShirokovVary2016prl,Gandolfi2017prc,LiMichelHu2019PRC,HiyamaLazauskas2016PRC,Fossez,Deltuva4n2018PL,
DeltuvaLazauskas2019PRC,DeltuvaLazauskas2019commentPRL,GandolfiHammer2019PRL,HiyamaKamimura2018FPhys}.

The experiment by Marqués et al. in 2002 lead to a number of theoretical investigations into whether a 4n state could exist \cite{Timofeyuk_2003,PhysRevLett.90.252501,Bertulani_2003}. These theoretical studies have shown that with current nuclear models, a 4n bound state could not exist. Only with a modification to the nuclear Hamiltonian, either via an additional four--body interaction \cite{Timofeyuk_2003} or other unrealistic enhancement of the nucleon--nucleon (NN) two--body interactions could a 4n bound state exist. Changes made to the well established nuclear Hamiltonians, such as the AV18/IL2 model in \cite{PhysRevLett.90.252501}, results to over binding in many light nuclei and even leads to a dineutron bound state.

Most theoretical studies that came out as a result of this 2002 experiment concluded that a 4n bound state or resonance could not exist with current nuclear theory. However, a more recent experiment by Kisamori et al. in 2016 shows evidence of a possible 4n resonance measured at an energy of $0.83\pm0.65(\mathrm{stat.})\pm1.25(\mathrm{sys.})~\mathrm{MeV}$ \cite{Kisamori} above the 4--body continuum. This most recent experiment sparked a renewed theoretical interest into this question \cite{ShirokovVary2016prl,Gandolfi2017prc,LiMichelHu2019PRC,HiyamaLazauskas2016PRC,Fossez,Deltuva4n2018PL,
DeltuvaLazauskas2019PRC,DeltuvaLazauskas2019commentPRL,GandolfiHammer2019PRL,HiyamaKamimura2018FPhys}. Some of these more recent theoretical studies agree with the earlier studies that a 4n bound or resonant state is unsupported by current theoretical models \cite{HiyamaLazauskas2016PRC,Fossez,Deltuva4n2018PL,
DeltuvaLazauskas2019PRC,DeltuvaLazauskas2019commentPRL,GandolfiHammer2019PRL,HiyamaKamimura2018FPhys}, while other studies confirm the possible existence of a 4n state as depicted in the 2016 experiment \cite{ShirokovVary2016prl,Gandolfi2017prc,LiMichelHu2019PRC}. This disagreement in whether a 4n state exists warrants further study of the low--energy scattering of few neutrons in the continuum, thus providing motivation for our most recent work published in Phys. Rev. Lett. \cite{PhysRevLett.125.052501}.

In this recently published work, we solve the trineutron (3n) and tetraneutron (4n) systems using the adiabatic hyperspherical framework, which has been successful at predicting resonances in few--body atomic systems in both a qualitative manner through the structure of the adiabatic potentials and quantitatively through an analysis of the phaseshift \cite{botero1986PRL,lin1975PRL}. These few neutron systems were studied primarily in the adiabatic approximation, neglecting non--adiabatic couplings to the excited states. The lowest adiabatic potential in both 3n and 4n systems are purely repulsive and qualitatively show no features to support a bound or resonant state above the scattering continuum. In addition, an elastic phaseshift analysis was performed for the lowest adiabatic potential with the second--derivative diagonal non--adiabatic correction that showed no features of a resonance at low energy, only an enhancement of the density of states caused largely by an attractive $\rho^{-3}$ long--range feature in the potential as it approaches the non--interacting limit. We attribute the low energy signal seen in the Kisamori et al. experiment to this enhancement, and not to a resonant 4n state \cite{PhysRevLett.125.052501}. The purpose of this article is to expand on this previous work to investigate the effects of non--adiabatic couplings to the excited states on this low--energy behavior in a more comprehensive multi--channel scattering treatment of these systems.

The rest of this article is organized in the following way. Section \ref{sec:level1} provides details on the adiabatic hyperspherical approach, including the numerical techniques used to compute the non--adiabatic couplings. The lowest few adiabatic hyperspherical potential curves for both 3n and 4n systems are reported in Sec. \ref{sec:Adiabatic_Potentials}, showing the repulsive nature of few interacting neutrons, with an emphasis on the lowest potential in both systems. In Sec. \ref{sec:non_adiacatic_couplings}, the non--adiabatic couplings are shown for the lowest few channels to provide qualitative and some quantitative features of the long--range behavior. Section \ref{sec:potentials_long_range} gives a detailed analysis of the long--range behavior of the lowest few adiabatic potentials, specifically providing the scattering length dependence of the $\rho^{-3}$ coefficient in these potentials at long--range. With knowledge of the long--range behavior of the adiabatic potentials and non--adiabatic couplings, a multi--channel treatment of the scattering of few neutron systems above the continuum is performed in Sec. \ref{sec:wigner_smith} and compared with the adiabatic treatment. Lastly, Sec. \ref{sec:conclusions} presents our conclusions. 

\section{\label{sec:level1}Adiabatic Hyperspherical Approach}
The 3n and 4n nuclear systems are explored in the framework of the hyperspherical representation. The main advantage of using this representation to study these few--body systems is that all of the dynamical features of the inter--particle interactions and reaction pathways at different length scales can be described both qualitatively and quantitatively on an equal footing through an adiabatic parameter denoted the hyperradius. Another key advantage of using the hyperspherical representation comes from its success in predicting resonances in few--body atomic systems. For example, photodetachment resonance in the positronium negative ion above the $n=2$ threshold was predicted by Botero and Greene in 1986 \cite{botero1986PRL} and confirmed by experiment in 2016 \cite{MichishioNagashima2016NatureComm}, and shape resonances in the $e-\mathrm{H}$ system by C.D.Lin in 1975 \cite{lin1975PRL} and confirmed by experiment in 1977 \cite{bryant1977PRL}. In both of these theoretical studies, the resonance features were observed both qualitatively in the structure of the adiabatic potential curves as well as quantitatively through an analysis of the elastic phaseshifts.

\subsection{Theoretical Formulation}
Within the adiabatic hyperspherical approach the 3n and 4n systems are solved using both an explicitly correlated Gaussian \cite{SV2,Rittenhouse_2011,RevModPhys.85.693,Varga1998} (CGHS) basis and the hyperspherical harmonic (HH) basis (\cite{Rittenhouse_2011} and references therein). The Hamiltonian for most systems can separate the center of mass coordinates with the relative coordinates, $\hat{H}=\hat{H}_{\mathrm{CM}}+\hat{H}_{\mathrm{rel.}}$. The center of mass Hamiltonian contains the kinetic energy operator of the center of mass. The Hamiltonian of the relative motion contains the relative hyperradial and hyperangular kinetic energy operators, as well as the potential energy.

The hyperangular kinetic energy and potential energy operators make up the adiabatic Hamiltonian, with the hyperradius treated initially as an adiabatic parameter. The generalized N--body adiabatic eigenvalue equation in hyperspherical coordinates to be solved is:

\begin{equation}
    \label{eq:supp_eigvalueprob}
    H_{ad}(\rho,\Omega)\Phi_\nu(\rho,\Omega)=U_\nu(\rho)\Phi_\nu(\rho,\Omega)
\end{equation}
where $\rho$ is the hyperradius, $\Omega$ is a set of hyperangles, and
\begin{multline}
    \label{eq:adexpression}
    H_{ad}(\rho,\Omega)=\frac{\hbar^2}{2\mu \rho^2}\biggr[\Lambda^2+\frac{(3N-4)(3N-6)}{4}\biggr]\\+V_{\mathrm{int.}}(\rho,\Omega)
\end{multline}
where $\mu_{\mathrm{N}}$ is the N--body hyperradial reduced mass.
The operator $\Lambda$ represents the hyperangular grand--angular momentum of the system and $V_{\mathrm{int.}}(\rho,\Omega)$ is the potential operator between the nucleons, containing two-body and three-body interaction terms.

The full N--body wavefunction in the relative coordinates is expanded in the eigenstates of Eq. \eqref{eq:supp_eigvalueprob}, giving the ansatz,
\begin{equation}
    \label{eq:ansatz}
    \Psi_{E}(\rho,\Omega)=\rho^{-\frac{3N-4}{2}}\sum_\nu F_{E,\nu}(\rho)\Phi_{\nu}(\rho,\Omega).
\end{equation}
The factor $(3N-4)(3N-6)/4$ in Eq. \eqref{eq:adexpression} comes from the multiplying factor of $\rho$ in Eq. \eqref{eq:ansatz}, which eliminates the first derivative in the hyperradial kinetic energy. Applying $\hat{H}_{\mathrm{rel.}}$ to Eq. \eqref{eq:ansatz}, projecting from the left with $\Phi_{\nu^{'}}(\rho,\Omega)$, and integrating over the hyperanglular coordinates and tracing over spin degrees of freedom leads to the following coupled hyperradial Schr\"odinger equations,

\begin{multline}
    \label{eq:coupled}
    \left(-\frac{\hbar^2}{2\mu}\frac{\partial^2}{\partial{\rho^2}}+W_{\nu}(\rho)-E\right)F_{\mathrm{E,\nu}}\left(\rho\right)\\
    -\frac{\hbar^2}{2\mu}\sum_{\nu^\prime\neq\nu}\left(2P_{\nu\nu^\prime}(\rho)\frac{\partial}{\partial{\rho}}+Q_{\nu\nu^\prime}(\rho)\right)F_{E,\nu^\prime}(\rho)=0
\end{multline}
where $P_{\nu \nu^{'}}(\rho)$ and $Q_{\nu \nu^{'}}(\rho)$ are first and second derivative non--adiabatic coupling matrix elements and $W_{\nu}(\rho)=U_{\nu}(\rho)-\frac{\hbar^{2}}{2\mu}Q_{\nu\nu}(\rho)$ is the $\nu^{\mathrm{th}}$ effective adiabatic potential \cite{Wang, e_+e_-Daily,DailyAsym}. 

Once Eq. \eqref{eq:supp_eigvalueprob} is solved for $U_{\nu}(\rho)$ and normalized fixed--$\rho$ eigenfunction $\Phi_{\nu}(\rho,\Omega)$, the next step is to calculate the first and second derivative non-adiabatic coupling matrix elements defined as \cite{Wang}:
\begin{equation}
    \label{eq:firstderivative}
    P_{\mu \nu}(\rho)=\biggr<\Phi_{\mathrm{\mu}}\biggr|\frac{\partial\Phi_{\nu}}{\partial \rho}\biggr>
\end{equation}

\begin{equation}
    \label{eq:secondderivative}
    Q_{\mu \nu}(\rho)=\biggr<\Phi_{\mathrm{\mu}}\biggr|\frac{\partial^2\Phi_{\nu}}{\partial \rho^2}\biggr>
\end{equation}
where the integrals are over the hyperangles. Symmetry properties of the $P$ matrix elements can be derived from manipulating Eq. \eqref{eq:firstderivative}. By differentiating the overlap $\langle\Phi_{\mathrm{\mu}}|\Phi_{\nu}\rangle$ with respect to $\rho$, it can be shown through Eq. \eqref{eq:firstderivative} that $P_{\mu\nu}(\rho)=-P_{\nu\mu}(\rho)$ and $P_{\nu\nu}(\rho)=0$. The diagonal second derivative coupling term added to the lowest adiabatic potential, $W_1(\rho)$, provides an upper bound to the ground-state, thus is important to include.

\subsection{\label{sec:nonadiabaticapproach}Non--Adiabatic Coupling: Numerical Approach}
The diagonal second derivative couplings can be re-written as $Q_{\nu\nu}(\rho)=-\langle\frac{\partial\Phi_{\mathrm{\nu}}}{\partial \rho}|\frac{\partial\Phi_{\nu}}{\partial \rho}\rangle$, thus one needs to only compute $\frac{\partial\Phi_{\nu}}{\partial \rho}$. One standard way to compute this derivative is to use finite-difference methods, however, we use a different approach that involves solving a matrix equation. The idea is to multiply Eq. \eqref{eq:supp_eigvalueprob} by $\rho^2$ then differentiate with respect to $\rho$. This leads to the following matrix equation \cite{PhysRevA.86.062511,Wang},

\begin{multline}
    \label{eq:eigvalueprob}
    \rho^2[U_\nu(\rho)-H_{ad}(\rho,\Omega)]\chi_\nu(\rho,\Omega)=\\
    [\frac{\partial }{\partial \rho}(\rho^2H_{ad}(\rho,\Omega))-\frac{\partial}{\partial \rho}(\rho^2U_{\nu}(\rho))]\Phi_\nu(\rho,\Omega)
\end{multline}
where,
\begin{equation}
    \label{eq:eigvalueprob1}
    \chi_\nu(\rho,\Omega)=\frac{\partial}{\partial \rho}\Phi_\nu(\rho,\Omega)+c\Phi_\nu(\rho,\Omega)
\end{equation}
and $c$ is solved for in an iterative process until the derivative only changes by a small amount, using the fact that $P_{\nu \nu}(\rho)=0$. 

Once the derivative of the channel functions are determined, the first--derivative coupling elements can be computed easily from Eq. \eqref{eq:firstderivative}. In general, the second--derivative of the channel functions are required to compute the second--derivative matrix elements, as indicated by Eq. \eqref{eq:secondderivative}. To avoid computing the second--derivative of the channel functions, the second--derivative couplings can be expressed in terms of the derivatives of the first--derivative matrix elements and channel functions through the relation \cite{Wang},
\begin{equation}
    \label{eq:non_adiabatic_relation}
    Q_{\mu \nu}(\rho)=\frac{\partial}{\partial\rho}P_{\mu \nu}(\rho)-\biggr<\frac{\partial\Phi_{\mathrm{\mu}}}{\partial\rho}\biggr|\frac{\partial\Phi_{\nu}}{\partial \rho}\biggr>.
\end{equation}
However, it has been shown that when solving the coupled hyperradial equations in Eq. \eqref{eq:coupled} by $R$--matrix propagation with the slow--variable discretization (SVD) method that only the component $-\langle\frac{\partial\Phi_{\mathrm{\mu}}}{\partial \rho}|\frac{\partial\Phi_{\nu}}{\partial \rho}\rangle$  of $Q_{\mu \nu}(\rho)$ is needed (see Appendix B of \cite{Wang}). Therefore, throughout the rest of this article, any further mention of second--derivative couplings refer to this component.

\section{Adiabatic Potentials}
\label{sec:Adiabatic_Potentials}
\begin{figure}[h!] 
\subfigure[]{\includegraphics[width=8.2 cm] {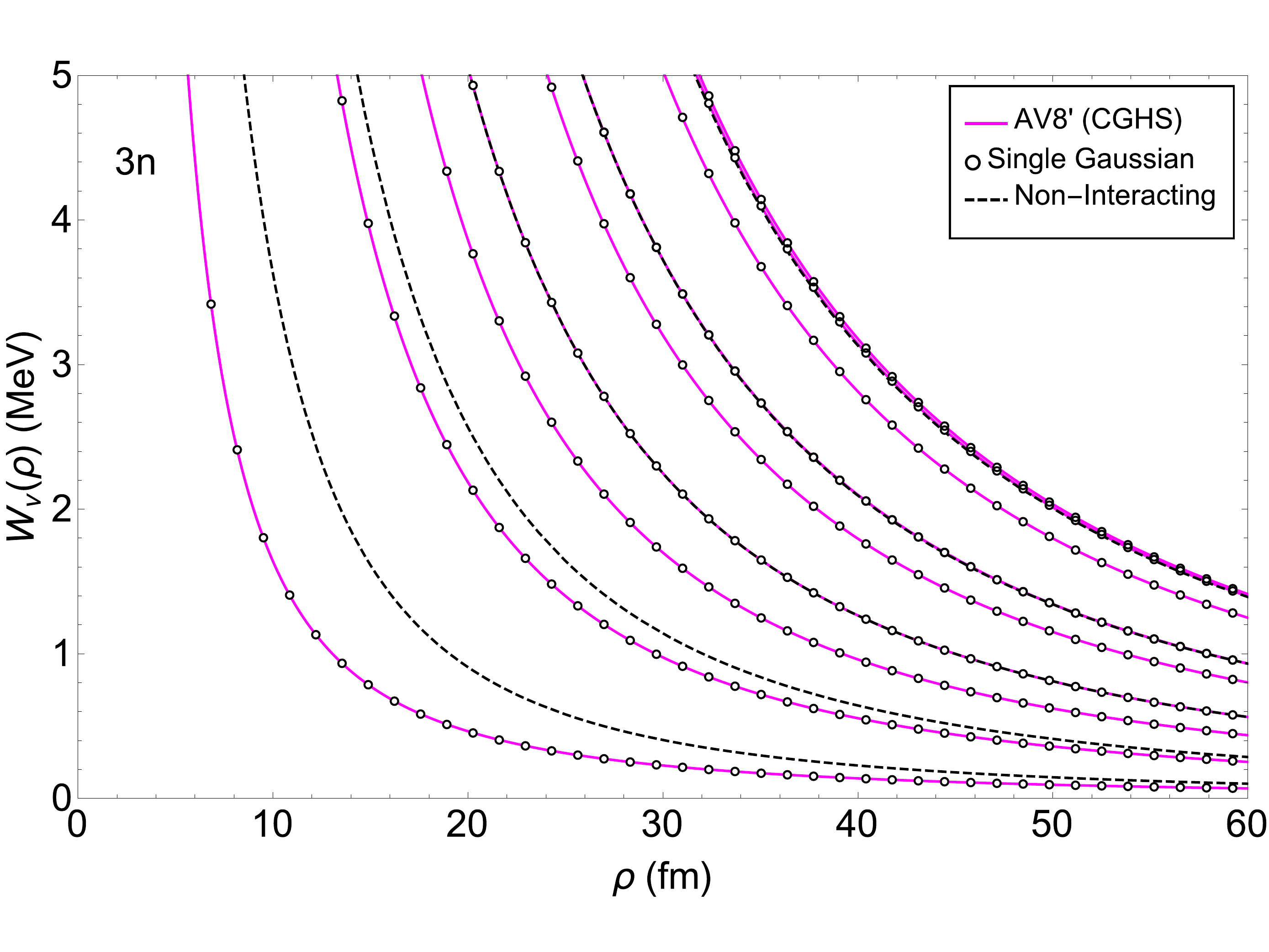}}

\subfigure[]{\includegraphics[width=8.2 cm] {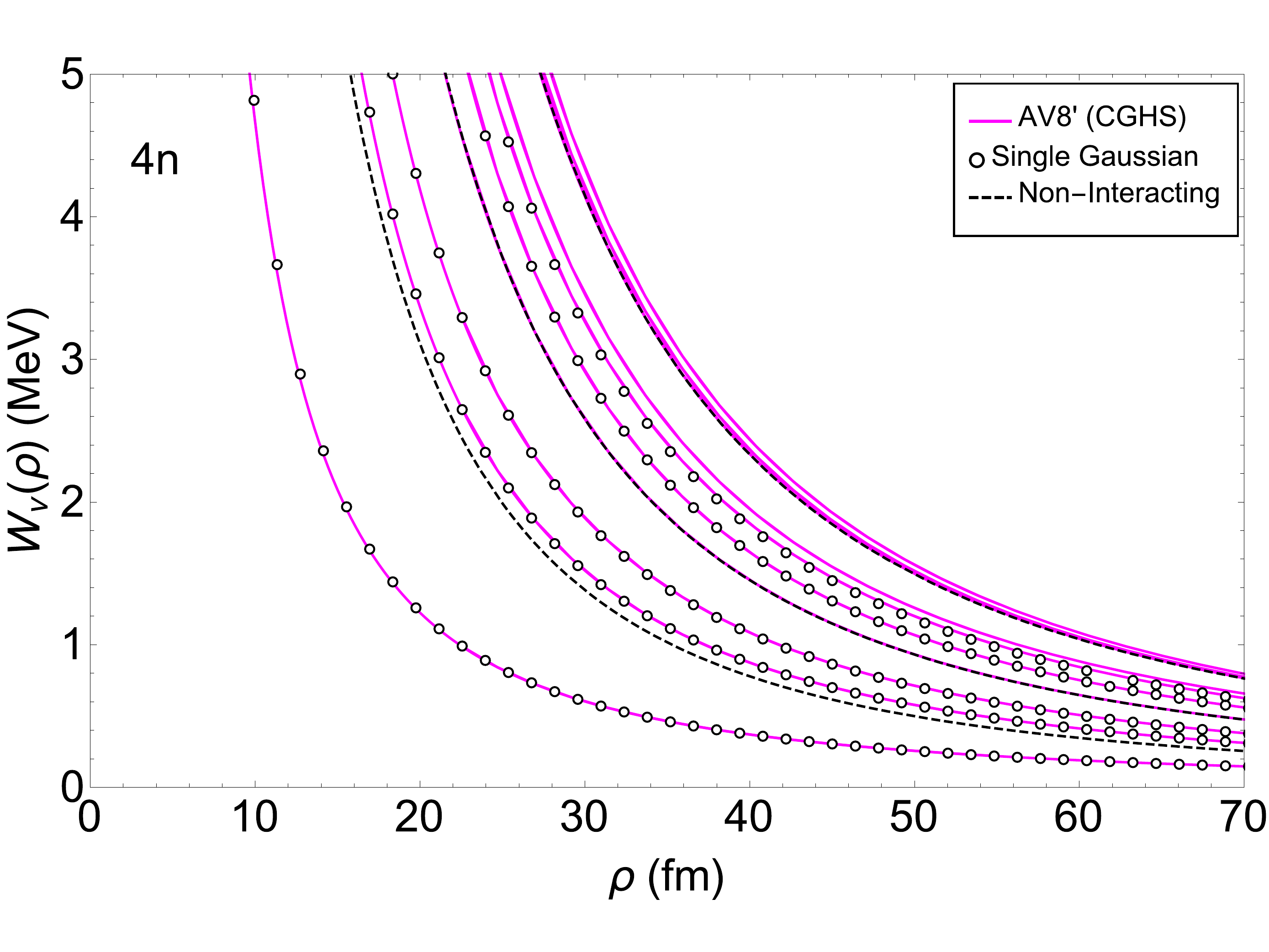}}
\caption{\label{fig:potentials} Adiabatic potentials for the 3n (a) and 4n (b) systems using the $\mathrm{AV8}^{\prime}$ two--body interaction. The diagonal second--derivative non--adiabatic couplings are included. The non--interacting potentials (dashed lines) are shown to provide a qualitative picture of the amount of attraction in these systems. Also, the adiabatic potentials are compared to a single--Gaussian model (open circles) with $L=0$ and $L=1$ natural parity states for the 4n and 3n systems, respectively.}
\end{figure}

The adiabatic hyperspherical potentials were obtained using two different basis expansions, the HH basis and the CGHS basis along with different nuclear models for the two--body interactions. The primary calculations were performed using the AV18 and $\mathrm{AV8}^{\prime}$ \cite{Wiringa:1994w,SV2} two--body nuclear potentials. Other nuclear interaction models are used with the HH basis to show a comparison of the adiabatic potentials among the various theories. The other NN interactions include the local NN potential model NVIa for the 3n system and models NVIa and NVIb for the 4n system, derived within the chiral effective field theory approach \cite{Piarulli:2016vel,Baroni:2018fdn}. The potentials were obtained for symmetries $J^{\pi}=0^{+}$ and $\frac{3}{2}^{-}$ for the 4n and 3n systems, respectively, providing the most attraction between the neutrons. The lowest few potentials, including the second--derivative non--adiabatic coupling term, for both 3n and 4n systems are shown in Figure \ref{fig:potentials} using the $\mathrm{AV8}^{\prime}$ interaction.

The lowest potential in both systems exhibit important features that aid in our understanding of the interaction between multiple neutrons. It is most evident from the potentials shown in Fig.~\ref{fig:potentials} that no bound or resonant states are supported for either the 3n or 4n system. There is no potential minimum or potential barrier that can support a bound or even temporarily bound resonant state that would indicate the existence of a trineutron or tetraneutron state. In fact, over the entire range of hyperradii, the potentials are purely repulsive due in part to the large Pauli repulsion in these fermionic systems. However, there is still significant attraction in these systems, which can be seen from a comparison with the non--interacting potential (dashed curves) in Fig. \ref{fig:potentials}. The lowest few adiabatic potentials lie well below the non--interacting potentials over a large range in the hyperradius, with the greatest attraction evident in the lowest potentials for both systems. 

\begin{figure}[!ht]
    \centering
    \subfigure[]{\includegraphics[width=8.2 cm]{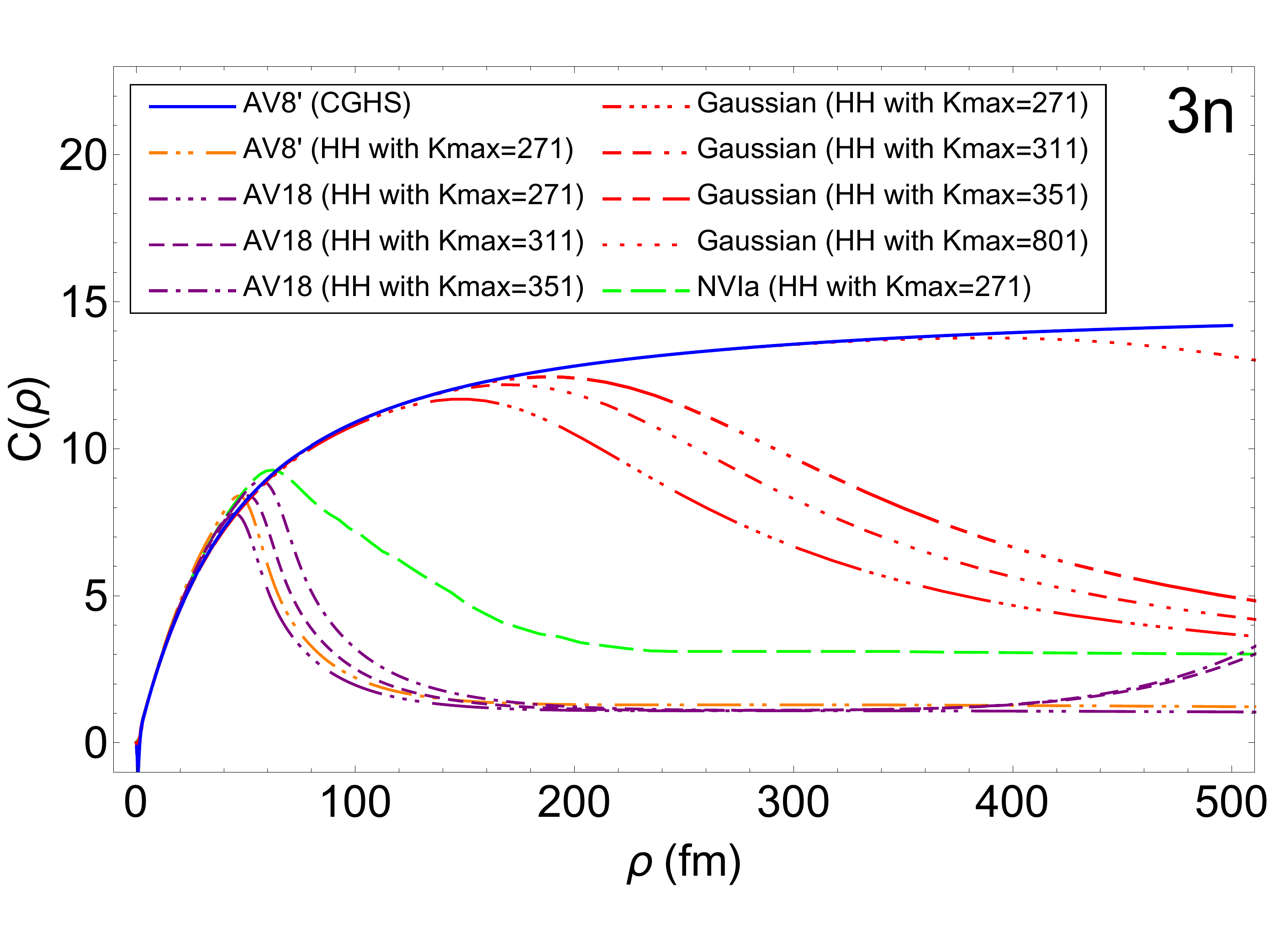}}

    \subfigure[]{\includegraphics[width=8.2 cm]{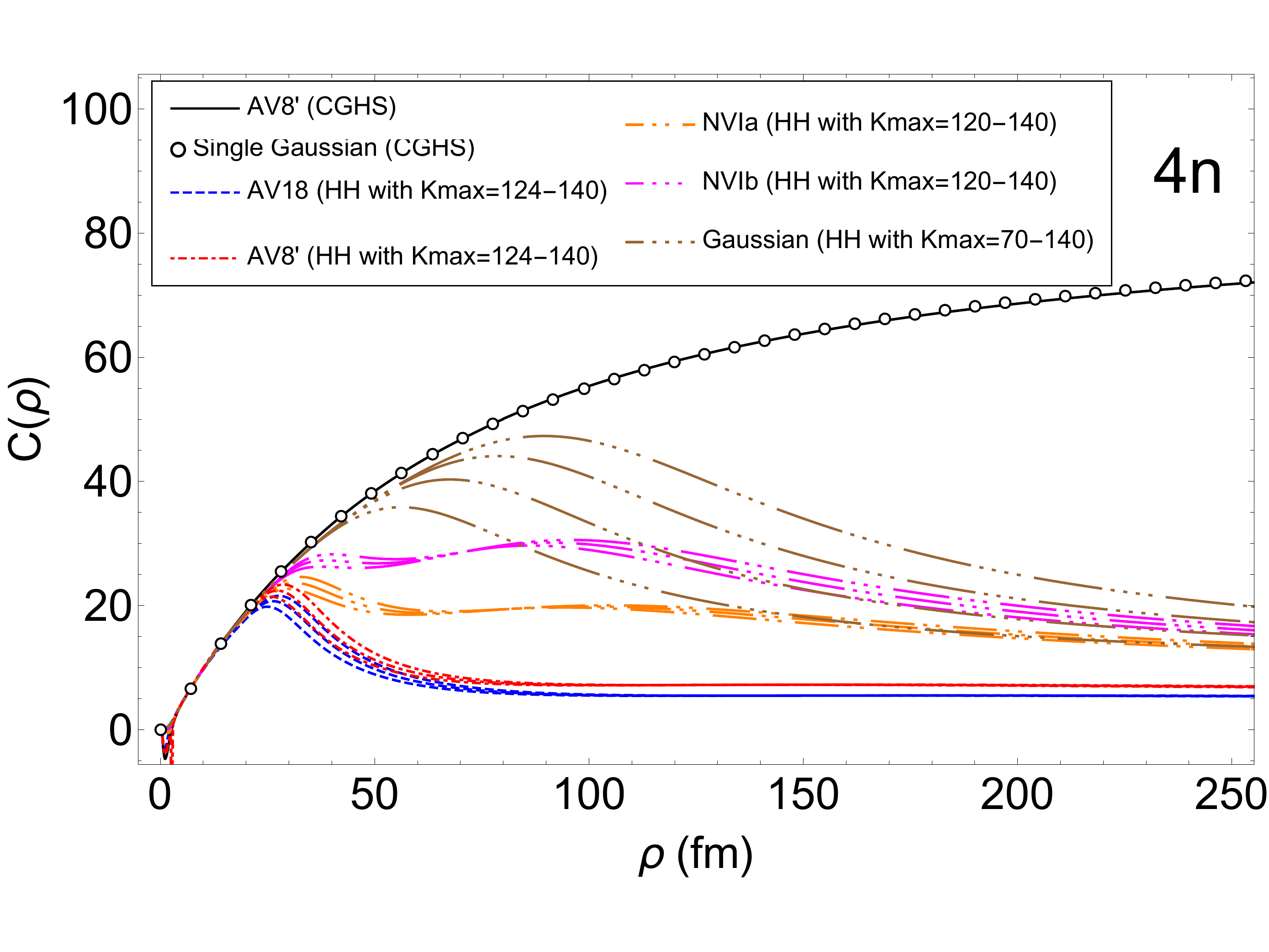}}
    
    \caption{\label{fig:Fig1}Plots of the dimensionless function $C_\nu(\rho) \equiv (\rho/a) [\rho^2 W_1(\rho) 2 \mu/\hbar^2 - l_{\rm eff}(l_{\rm eff}+1) ]$ for the 3n (a) and 4n (b) systems with $\nu=1$. According to Eq. \eqref{eq:asymform}, we should obtain $C_\nu(\rho \rightarrow \infty)=C_\nu$, where $C_\nu$ is the coefficient listed in Table \ref{table:C3_coeff}. We observe the slow convergence for large $\rho$ of the adiabatic potentials calculated using the HH basis. However, it has to be noted that where the convergence is achieved, the functions $C_\nu(\rho)$ obtained for the different  interactions used in this work almost collapse onto a single curve. Noticeably, this happens already for fairly small values of rho, showing that the adiabatic potentials are already universal at moderate values of the hyperradius. In fact, the limit $C_\nu(\rho)=C_\nu$ is reached only for $\rho>500\ {\rm fm}$.  
    }
\end{figure}

The long--range behavior of the adiabatic potentials plays an important role in determining the low--energy behavior of the scattering phaseshift. From Fig. \ref{fig:potentials}, the lowest few adiabatic potentials in both neutron systems approach the non--interacting potentials as $\rho^{-3}$. It is found that the asymptotic form of the lowest few potentials take the form,
\begin{equation}
    \label{eq:asymform}
    W_{\nu}(\rho)\rightarrow\frac{\hbar^2}{2\mu}\biggr(\frac{l_{\mathrm{eff}}(l_{\mathrm{eff}}+1)}{\rho^2}+\frac{C_{3,\nu}}{\rho^3}\biggr)
\end{equation}
where $l_{\mathrm{eff}}$ is the effective angular momentum quantum number, and where $C_{3,\nu}=C_{\nu}a$, with $a$ being the $s$--wave scattering length, and $C_{\nu}$ is a dimensionless constant that depends on system size, symmetry and eigenstate. For the lowest adiabatic potential, the $\rho^{-3}$ behavior is emphasised through a plot, shown in Figure \ref{fig:Fig1}, of $C(\rho)=(\rho/a)((2\mu/\hbar^{2})W_{1}(\rho)-l_{\mathrm{eff.}}(l_{\mathrm{eff.}}+1))$ with $l_{\mathrm{eff.}}=30$ for the lowest potential of the 4n system and $l_{\mathrm{eff.}}=5/2$ for the lowest 3n potential. These have been re-scaled by the spin--singlet $s$--wave scattering length. Figure \ref{fig:Fig1} shows a comparison of $C(\rho)$ for both neutron systems obtained for different nuclear interaction models and basis sets. In particular, these figures show the slow convergence of the HH basis at large values of the hyperradius compared to calculations performed using the CGHS basis, which provides the best converged results beyond 30--50 fm. At small hyperradii, both basis sets are well converged and the potential curves computed with different nuclear interaction models are nearly indistinguishable. Only at large hyperradii is there a clear difference in the lowest potential, which is largely due to the slow convergence of the HH basis.

Universal behavior is observed in these few--neutron systems, as shown by the remarkable agreement between the adiabatic potentials in Fig.~\ref{fig:potentials} using both the accurate and well established $\mathrm{AV8}^{\prime}$ nuclear interaction and a single Gaussian potential, whose parameters are given in Table~\ref{table:gauss_params}. This agreement resulting from two qualitatively different interactions demonstrates that interacting neutrons exhibit long--range universal physics, further supported by similarities in the elastic phaseshifts \cite{PhysRevLett.125.052501}. This can be understood due to the large ratio of the singlet $s$--wave scattering length to the range of the nuclear interaction, which is approximately $|a_s/r_0|\approx10$. The connection between short--range interactions with large scattering lengths to universal physics has been extensively studied in various contexts relating to not only the Efimov effect in atomic and nuclear three--body systems \cite{DIncaoReview,Rittenhouse_2011,Greene2017RMP,NaidonReview,PhysRevA.78.062701,PhysRevLett.111.132501,Efimov1988,KievskyandG2016,PhysRevC.100.034004,PhysRevC.95.024001,PhysRevLett.121.072701}, but also in connection to BCS--BEC crossover in few--fermion atomic systems \cite{Rittenhouse_2011,regal2007,PhysRevLett.99.090402}. Furthermore, the 3n and 4n adiabatic potentials using different and well established nucleon--nucleon interactions are universal at moderately small hyperradii (from 5--30 fm), with qualitative differences in the range $0<\rho<5$~fm, as seen in Fig. \ref{fig:Fig1}. This universal behavior in these neutron few--body systems is explored further in Sec.~\ref{sec:potentials_long_range}, analyzing the long--range behavior of these adiabatic potentials at different scattering lengths up to unitarity.


\begin{figure}[h!] 
\hspace{-0.0in}\includegraphics[width=8.2 cm] {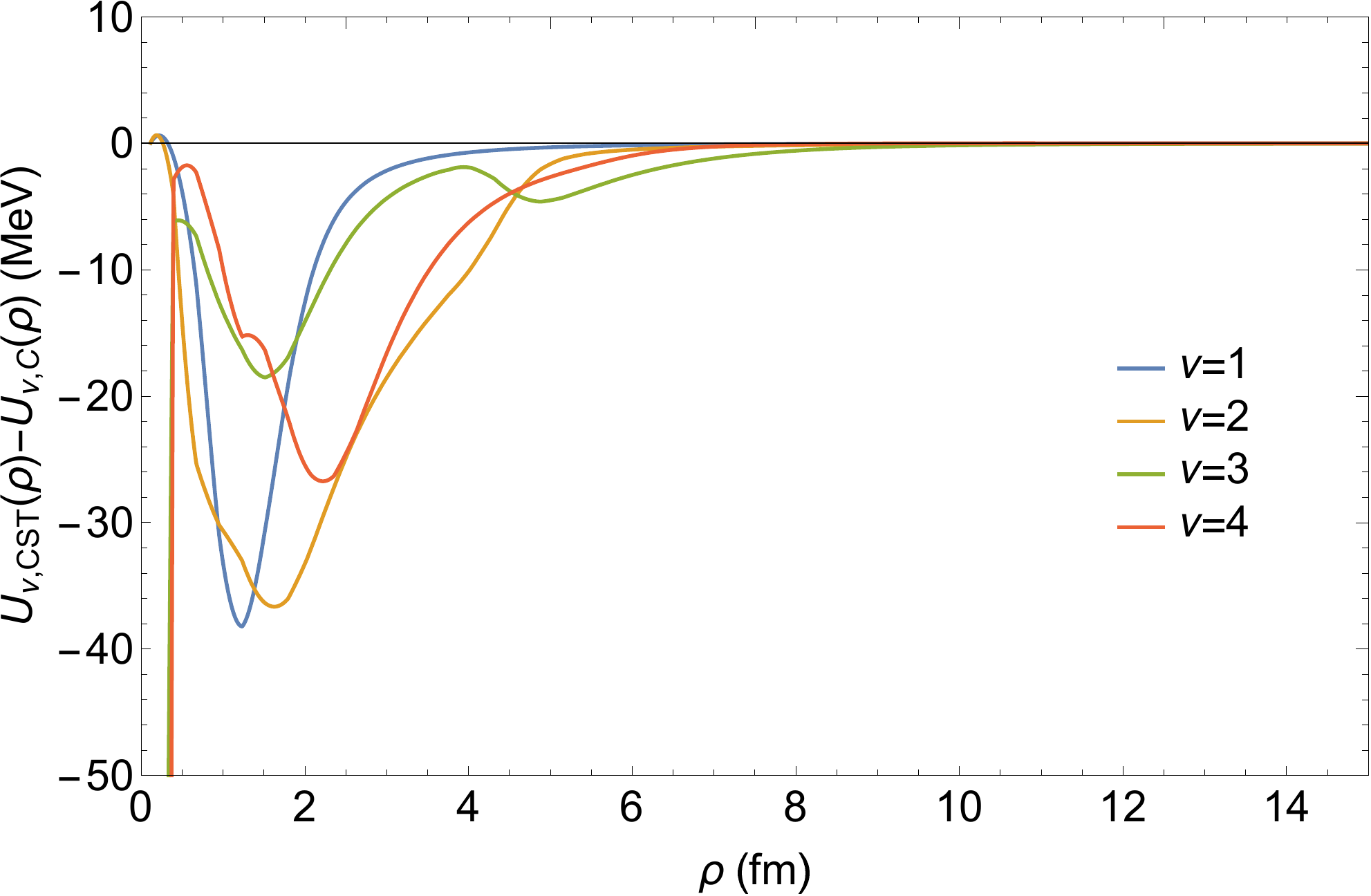}
\caption{\label{fig:4nPotentail_Difference} Difference in the lowest few 4n hyperradial potentials for $L=0$ states ($\nu=1-4$) with and without the spin--orbit and tensor interactions. The differences suggest that the tensor and spin--orbit interactions provide attraction to the system at hyperradii around $0<\rho<5$~fm.}
\end{figure}

One key interest concerning the neutron--neutron interaction is the role of tensor and spin--orbit interactions for systems of few neutrons. In nucleon--nucleon interactions, the tensor and spin--orbit interactions are important for binding the neutron and proton in the spin--triplet and isospin--singlet state \cite{Wiringa:1994w}. For neutron systems on the other hand, the tensor and spin--orbit interactions do not lead to binding of two neutrons. However, these interactions do provide extra attraction at short--range ($0<\rho<4~\mathrm{fm}$). This is best illustrated in Fig.~\ref{fig:4nPotentail_Difference} for the 4n system, showing the difference in the lowest few adiabatic potentials for $L=0$ states with and without the tensor and spin--orbit interactions using the $\mathrm{AV8}^{\prime}$ nuclear interaction model.

\begin{figure}[!ht]
    \centering
    \subfigure[]{\includegraphics[width=8.2 cm]{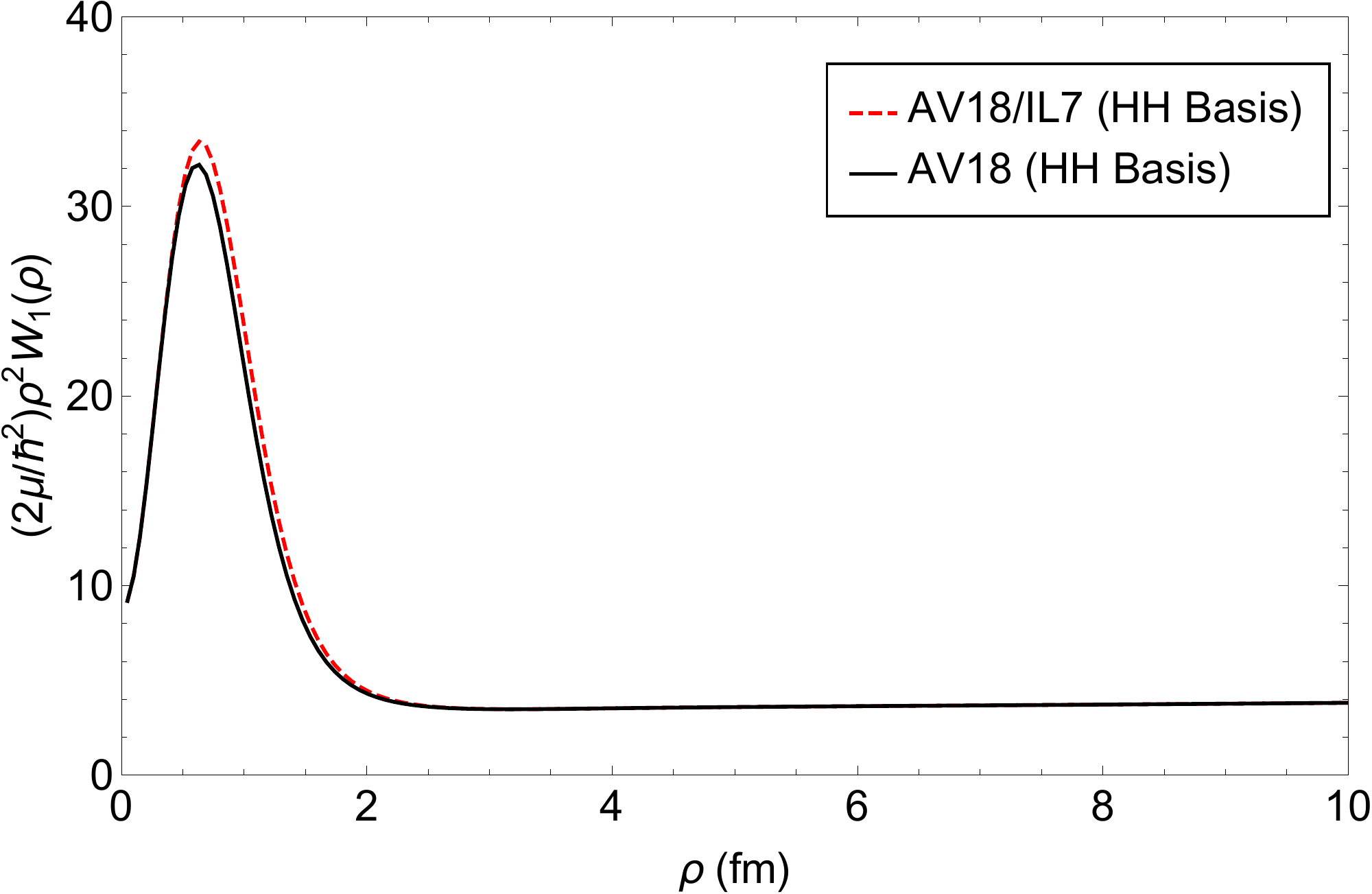}}

    \subfigure[]{\includegraphics[width=8.2 cm]{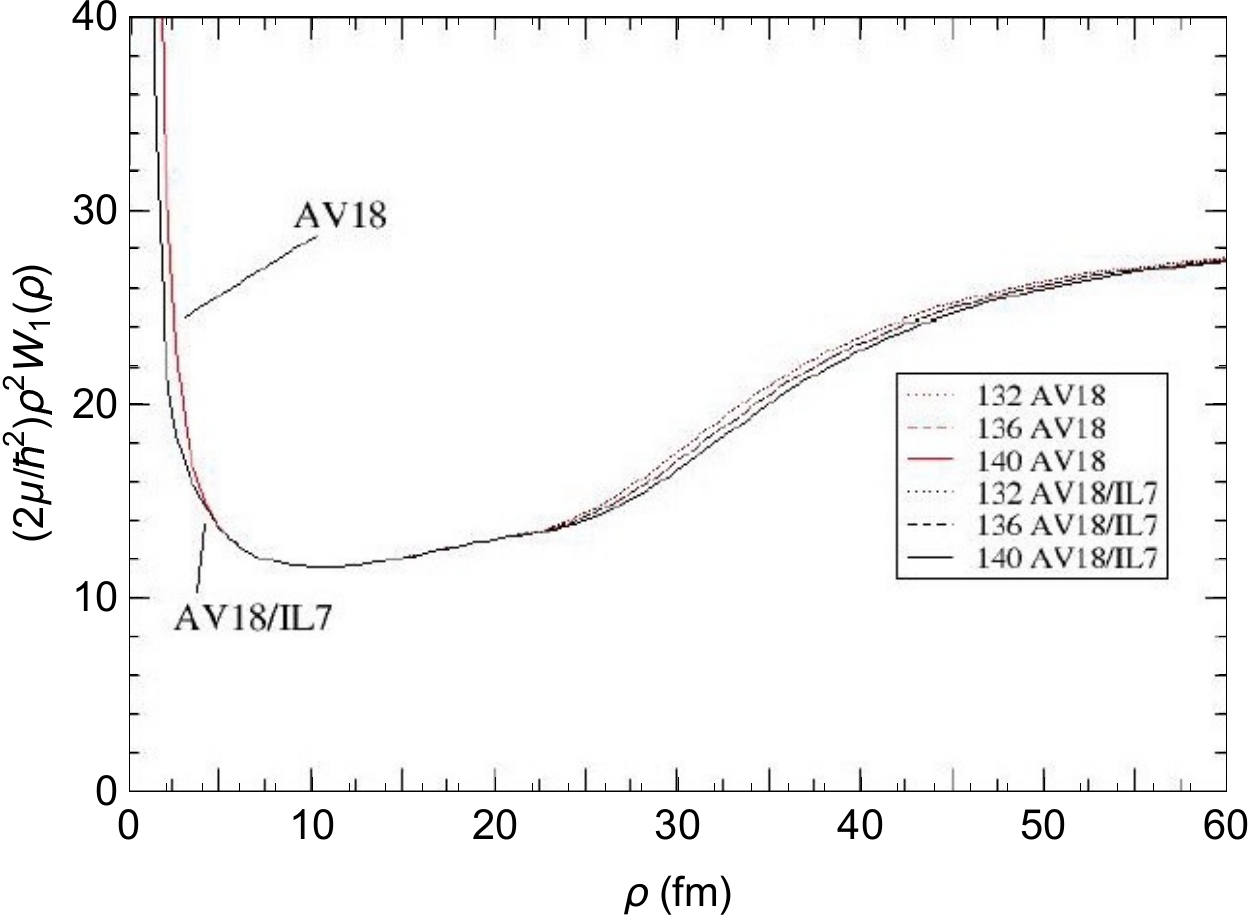}}
    
    \caption{\label{fig:3N_Force} A comparison of the lowest adiabatic potential using the AV18 two--body interaction with and without the IL7 3N force for the 3n (a) and 4n (b) systems. There is only a significant difference for $\rho <$ 5 fm.}
\end{figure}

In many nuclear systems, the 3N force play a significant role in determining the correct binding energy of light nuclei \cite{UIX,UIX2,Illinois3BF,PieperIL7,PhysRevLett.120.052503}. Thus, a logical question would be whether or not the 3N force plays a significant role in binding of few--neutron systems. To address this question, the 3n and 4n systems were studied using the AV18 two--body potential with the Illinois 7 (IL7) model of the 3--body force. A comparison between the lowest potential in both 3n and 4n systems with and without the IL7 3N term is shown in Fig. \ref{fig:3N_Force}. As shown in Fig. \ref{fig:3N_Force}, the lowest re--scaled potentials for these systems is represented by the quantity, $(2\mu/\hbar^{2})\rho^{2}U(\rho)$, which approaches 30 at large $\rho$ for the 4n system and 8.75 for the 3n system. From the comparison of this potential with and without the IL7 3N force, there is only a significant difference for $\rho<5$~fm, with the 3N force adding slight attraction to the 4n system and slight repulsion in the 3n system. At large values of $\rho$, which governs the low--energy scattering of the system, the potentials are nearly identical. Thus, it is concluded the 3N force plays little role in the low--energy regime for 3n and 4n scattering with providing not enough attraction to lead to a bound or even resonant bound state.

\section{Non-adiabatic Coupling Matrix Elements}
\label{sec:non_adiacatic_couplings}

\begin{figure}[!ht]
    \centering
    \subfigure[]{\includegraphics[width=8.2 cm]{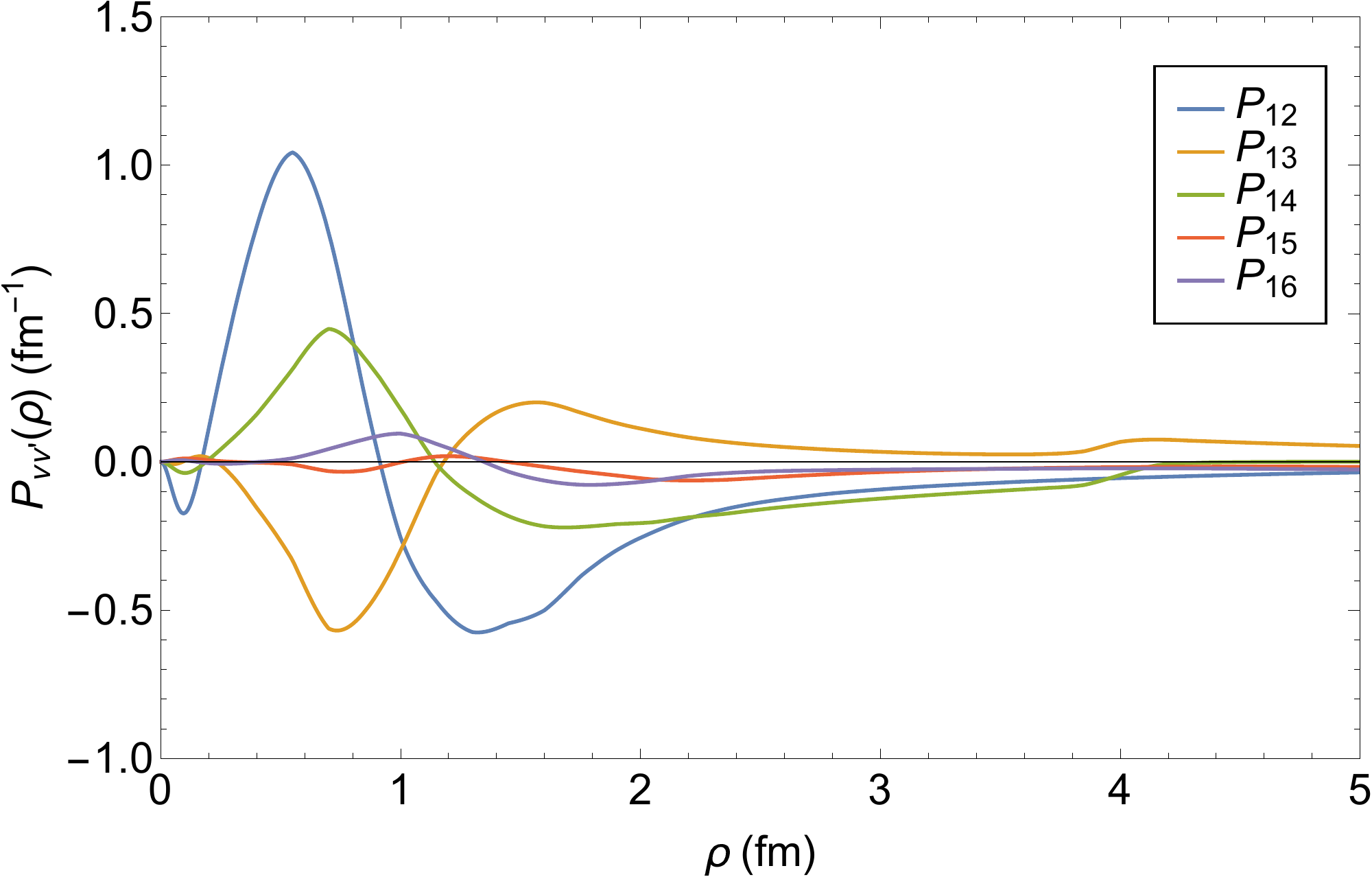}}

    \subfigure[]{\includegraphics[width=8.2 cm]{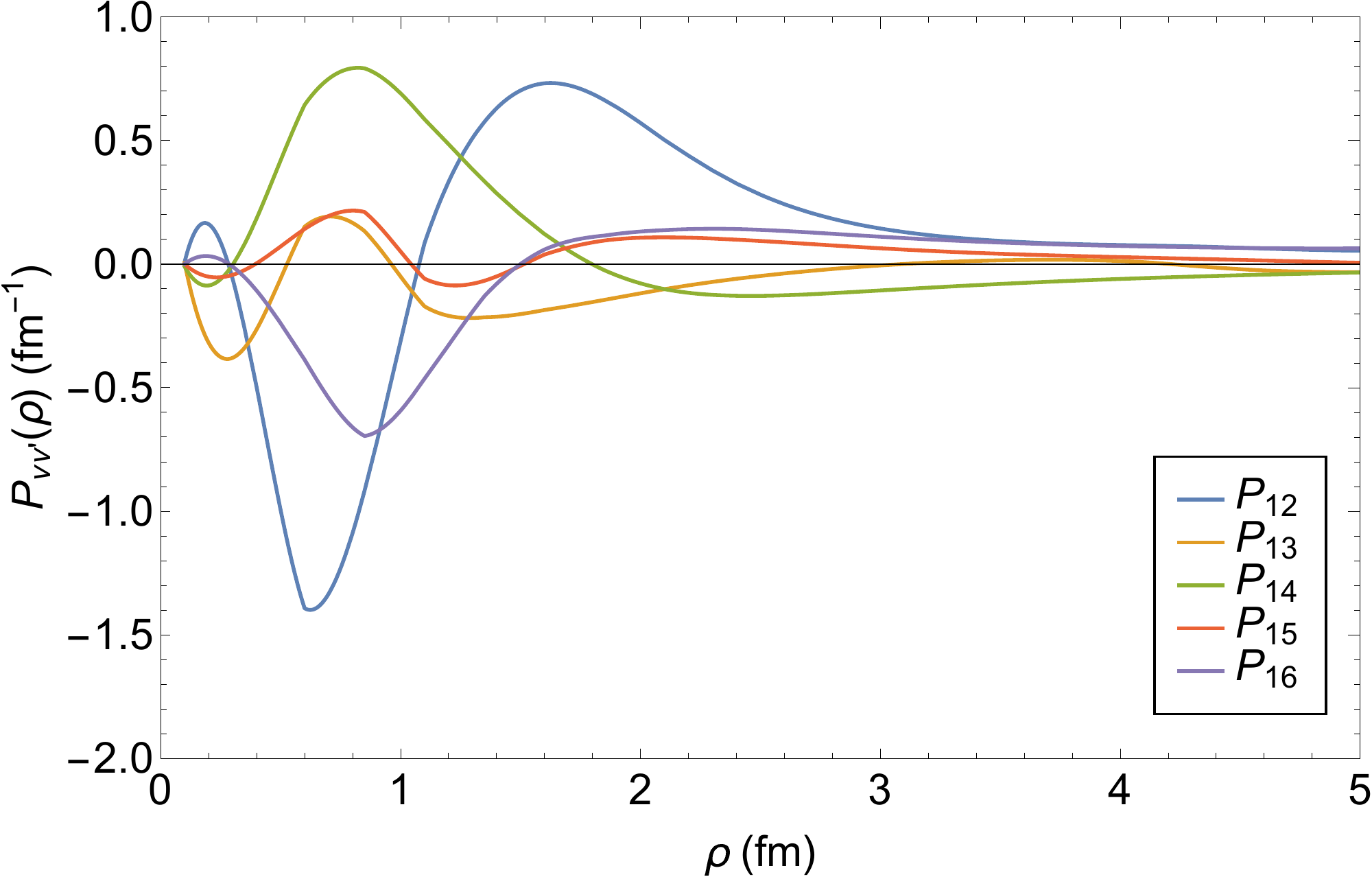}}
    
    \caption{The first--derivative coupling matrix elements $P_{1\nu}(\rho)$ in the lowest channel for the 3n (a) and 4n (b) systems with $\nu=2-6$ (from Eq. \eqref{eq:firstderivative} $P_{11}(\rho)=0$).}
    \label{fig:4n_Pmat}
\end{figure}
\begin{figure}[!ht]
    \centering
    \subfigure[]{\includegraphics[width=8.2cm]{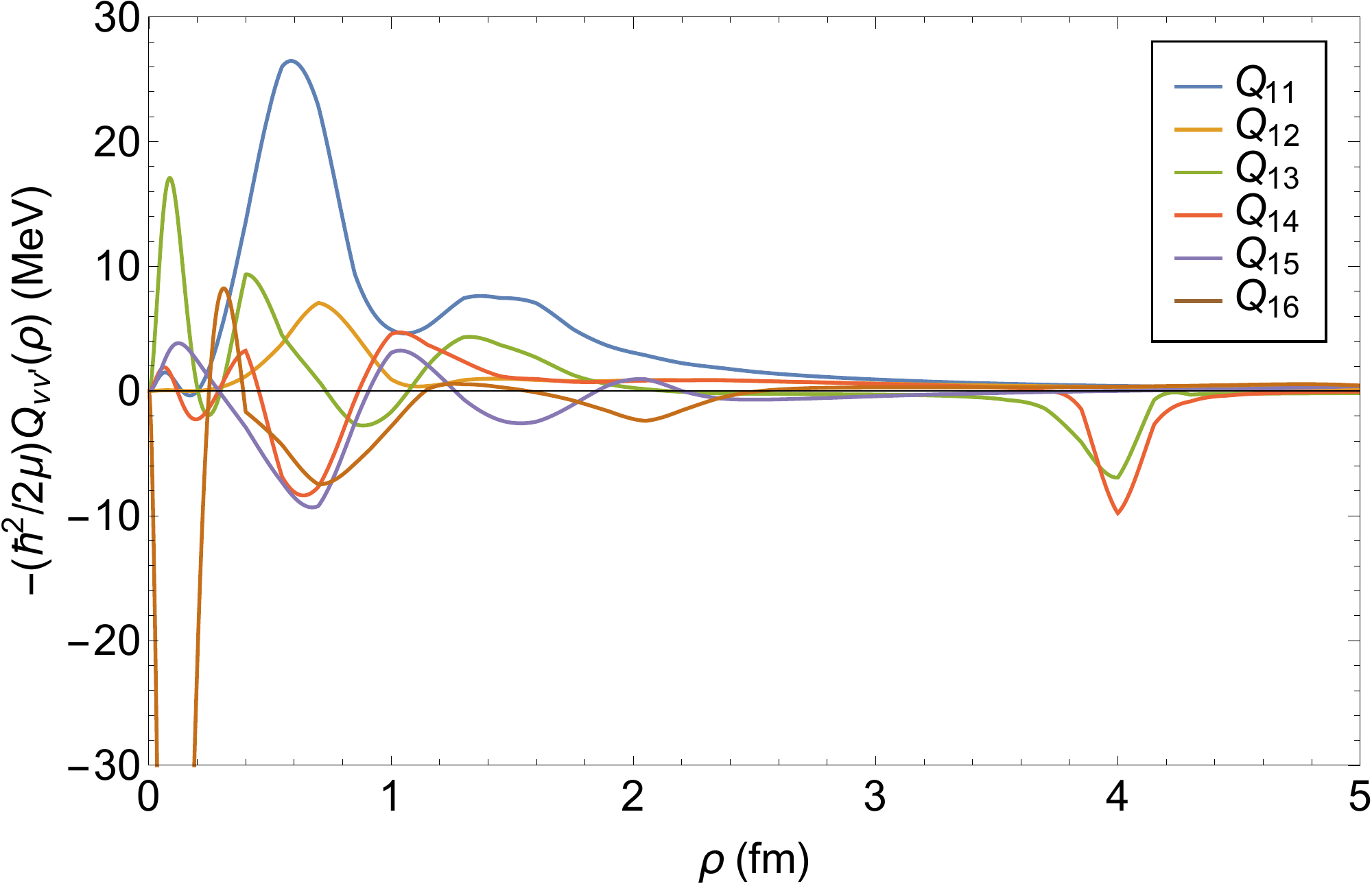}}

    \subfigure[]{\includegraphics[width=8.2 cm]{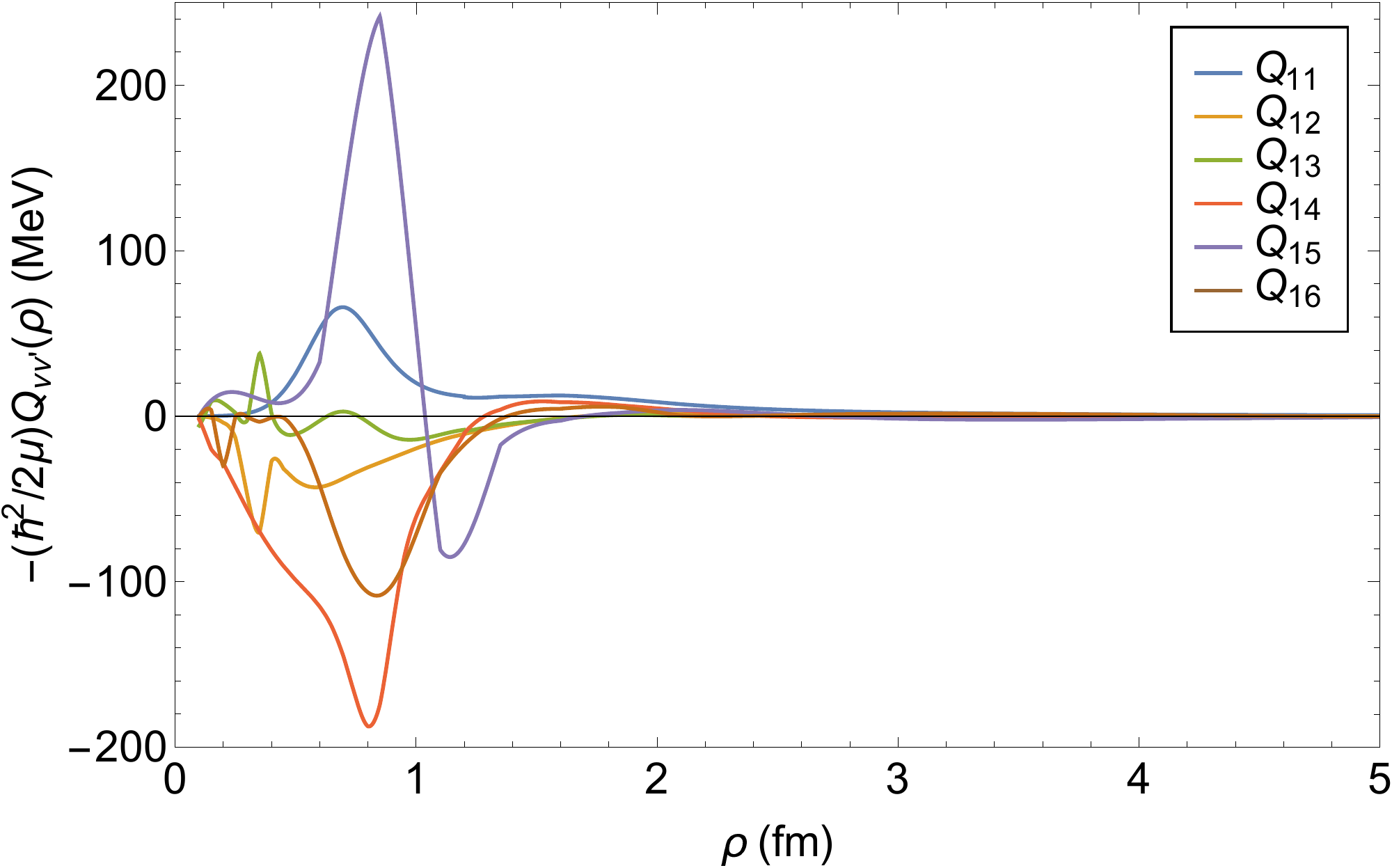}}
    
    \caption{The second--derivative coupling matrix elements $Q_{1\nu}$ in the lowest channel for the 3n (a) and 4n (b) systems with $\nu=1-6$. The coupling matrix elements are scaled by $\hbar^{2}/2\mu$.}
    \label{fig:4n_Qmat}
\end{figure}

 Before considering the full treatment of the coupled hyperradial Schr$\ddot{\mathrm{o}}$dinger equation, the first-- and second--derivative non--adiabatic coupling matrix elements need to be computed at each hyperradius. The non--adiabatic matrices, defined in Eqs. \eqref{eq:firstderivative} and \eqref{eq:secondderivative}, are computed using the methods described in section \ref{sec:nonadiabaticapproach}. The long--range behavior of these matrix elements are analyzed with an emphasis on the non--adiabatic couplings for the 3n system, noting the asymptotic behavior is equivalent in the 4n system. Through this study, it is determined that the second--derivative coupling matrix elements fall off faster than $\rho^{-3}$, which do not contribute to the long--range potential given in Eq. \eqref{eq:asymform}. The non--adiabatic couplings between the lowest channel and lowest 6 channels are shown in Figures \ref{fig:4n_Pmat} and \ref{fig:4n_Qmat} for the first-- and second--derivative elements, respectively.

\begin{figure}[h!] 
\hspace{-0.0in}\includegraphics[width=8.2 cm] {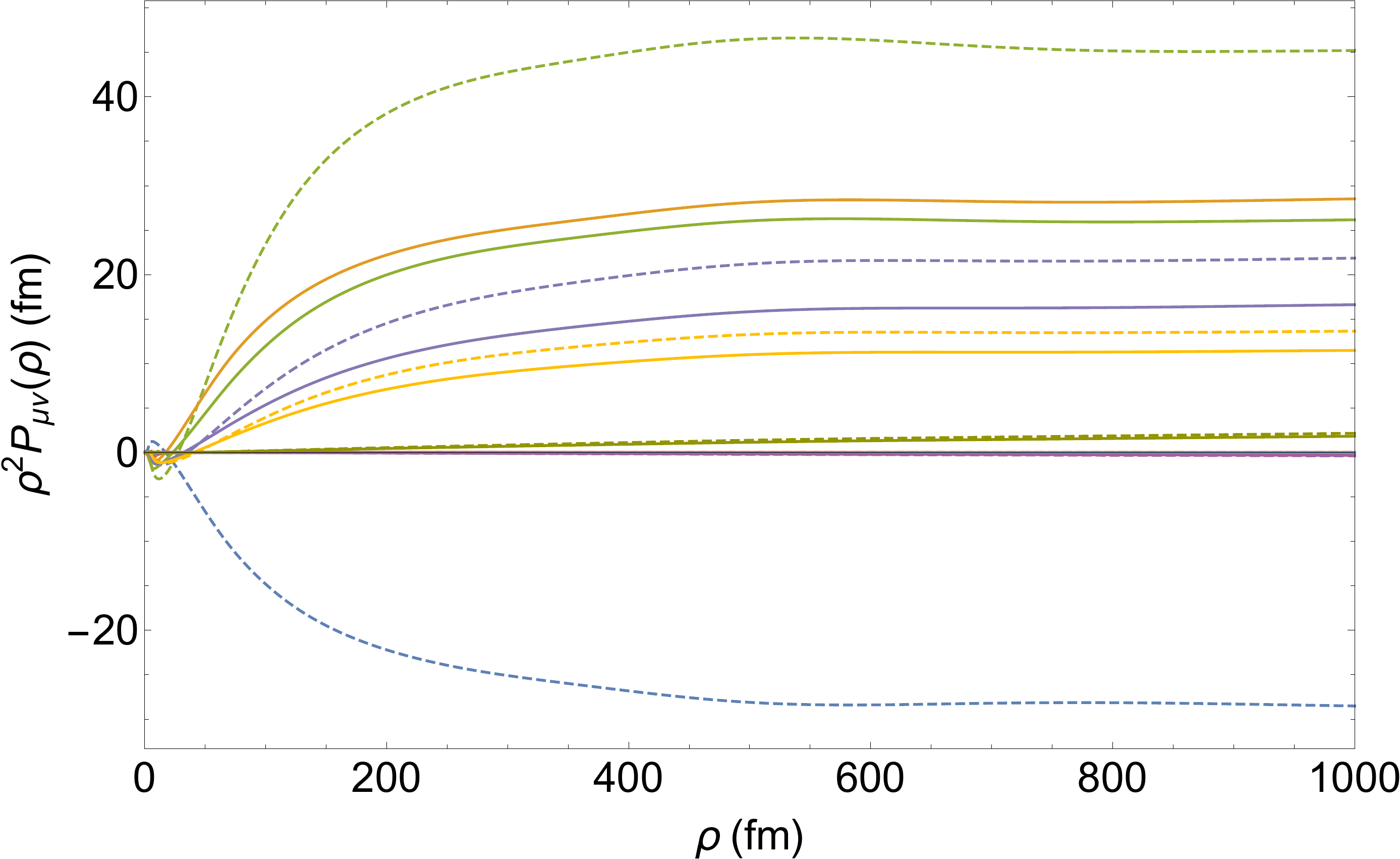}
\caption{\label{fig:3n_DiagQ} The first--derivative non--adiabatic matrix elements for the lowest few 3n channels are shown multiplied by $\rho^2$ to emphasise the long--range behavior. The solid lines are the coupling matrices to the first channel and the dashed lines are coupling matrices to the second channel (i.e. $P_{1\nu}$ and $P_{2\nu}$ with $\nu=1,..,6$).}
\end{figure}

\begin{figure}[h!] 
\hspace{-0.0in}\includegraphics[width=8.2 cm] {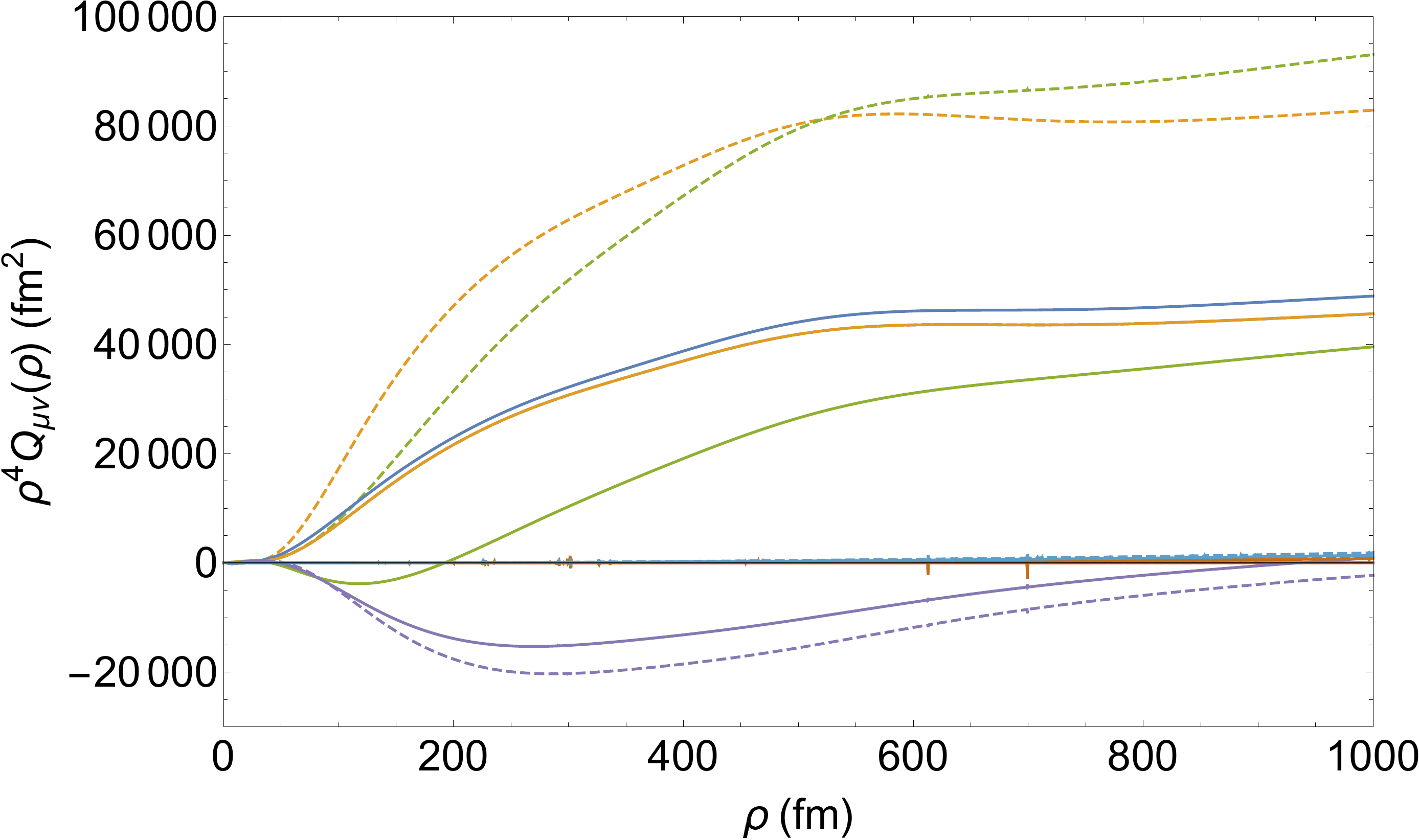}
\caption{\label{fig:4n_DiagQ} The second--derivative non--adiabatic matrix elements for the lowest few 3n channels are shown multiplied by $\rho^4$ to emphasise the long--range behavior. The solid lines are the coupling matrices to the first channel and the dashed lines are coupling matrices to the second channel (i.e. $Q_{1\nu}$ and $Q_{2\nu}$ with $\nu=1,..,6$).}
\end{figure}

The first-- and second--derivative couplings for the 3n and 4n systems only have a significant impact on the adiabatic potentials over a range of the hyperradius from $0\leq\rho\leq5~\mathrm{fm}$, beyond which they rapidly decrease to zero. The long--range behavior of these non--adiabatic matrix elements can be seen by multiplying $P_{\mu\nu}(\rho)$ by $\rho^2$ and $Q_{\mu\nu}(\rho)$ by $\rho^4$, which is shown in figures \ref{fig:3n_DiagQ} and \ref{fig:4n_DiagQ} for the first-- and second--derivative matrix elements for the 3n system, respectively. From these figures, it is evident the second--derivative matrix elements are short--range, falling off faster than $\rho^{-3}$, in--fact falling off as $\rho^{-4}$. The first--derivative matrix elements exhibit long--range behavior, falling off like $\rho^{-2}$ at large hyperradii. Likewise, this long--range behavior in the non--adiabatic couplings is also observed in the 4n system.

This analysis of the long--range behavior of the non--adiabatic matrix elements indicates that the diagonal second--derivative matrix elements do not impact the long--range behavior of the adiabatic potentials, although the short--range behavior is affected. As a result, when solving the coupled hyperradial equation, the couplings only have significant influence on the collision eigenphaseshifts at high scattering energies, while the low--energy behavior is only slightly modified due to the long--range behavior of the first--derivative couplings. Section \ref{sec:wigner_smith} discusses the affect of the non--adiabatic coupling terms on the energy--dependent eigenphaseshifts.

\section{Long-Range Behavior of the Adiabatic Potentials}
\label{sec:potentials_long_range}
In these few neutron systems, it is observed that the lowest few adiabatic potentials take the form of Eq. \eqref{eq:asymform}, in which the the potentials deviate from the non--interacting limit by a $\rho^{-3}$ dependence that scales linearly in the two--body $s$--wave scattering length. This dependence of the asymptotic form on the two--body $s$--wave scattering length has been shown for hyperspherical potentials associated with the many particle continuum, for systems with short--range interactions having a two--body scattering length larger than the range of the interaction \cite{PhysRevA.38.1193,PhysRevA.60.1451}. The asymptotic form is valid with and without the diagonal second derivative coupling term since the coupling terms fall off as $\rho^{-4}$, as was shown in Sec. \ref{sec:non_adiacatic_couplings}.

\begin{figure}[!ht]
    \centering
    \includegraphics[width=8.5cm]{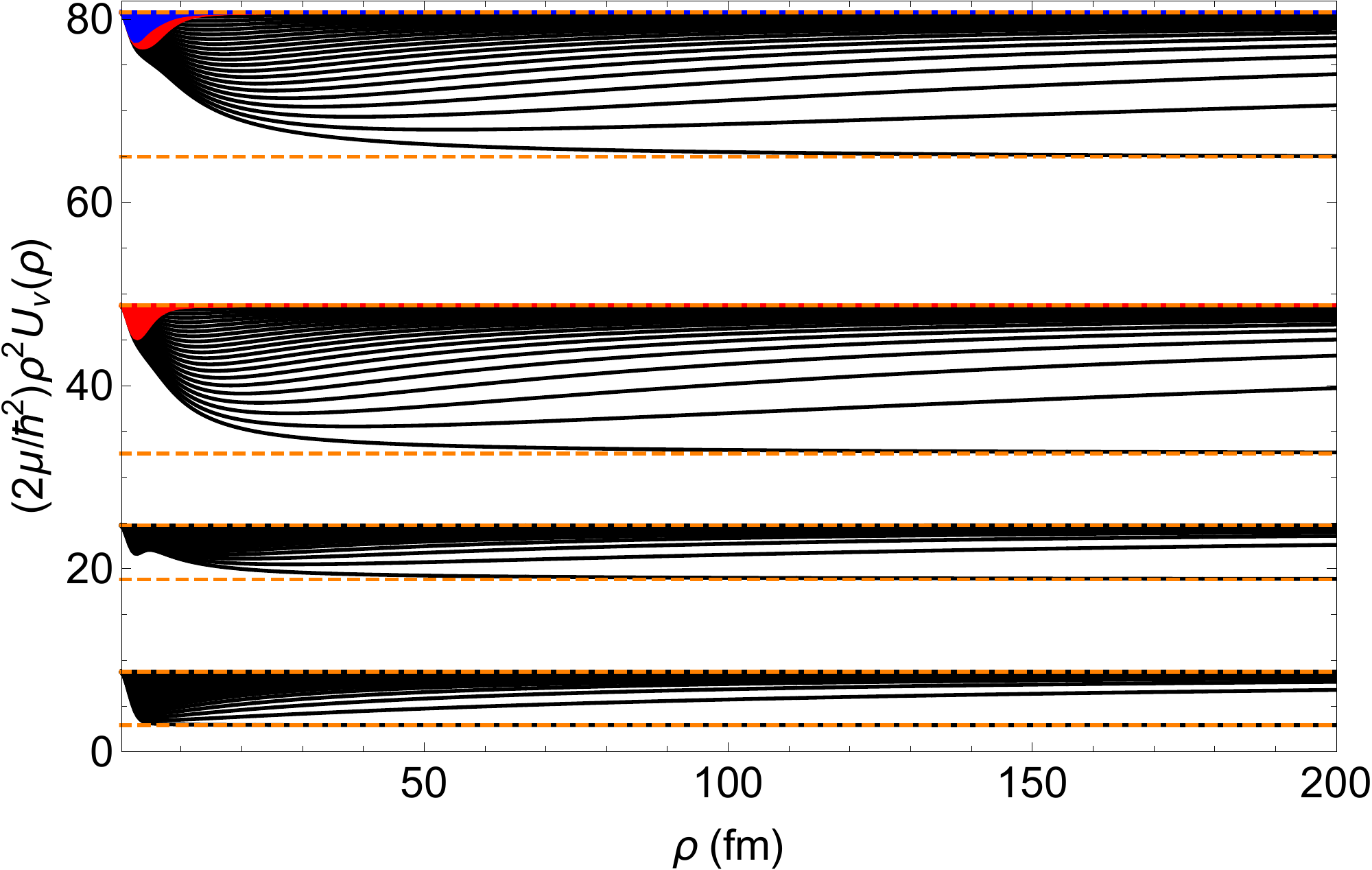}
    \caption{The lowest 7 adiabatic potentials for the 3n system for the $J^{\pi}=\frac{3}{2}^{-}$ symmetry with $L=1$ using a single--Gaussian potential with $r_{0}=1.78$~fm. The strength $V_0$ is tuned to give different scattering lengths from 0 to infinity with no two--body bound state. Each set of potentials represent a different $s$--wave scattering length in this range with limits shown by the horizontal dashed lines representing the non--interacting limit (upper--dashed) and unitarity limit (lower--dashed). The unitarity limits are given in Table \ref{table:C3_coeff}. The black curves represent the potentials with the $\rho^{-3}$ behavior as they approach the non--interacting limit and the red and blue curves represent those that fall off faster than $\rho^{-3}$.}
    \label{fig:C3_Coeff3n}
\end{figure}

\begin{figure}[!ht]
    \centering
    \includegraphics[width=8.5cm]{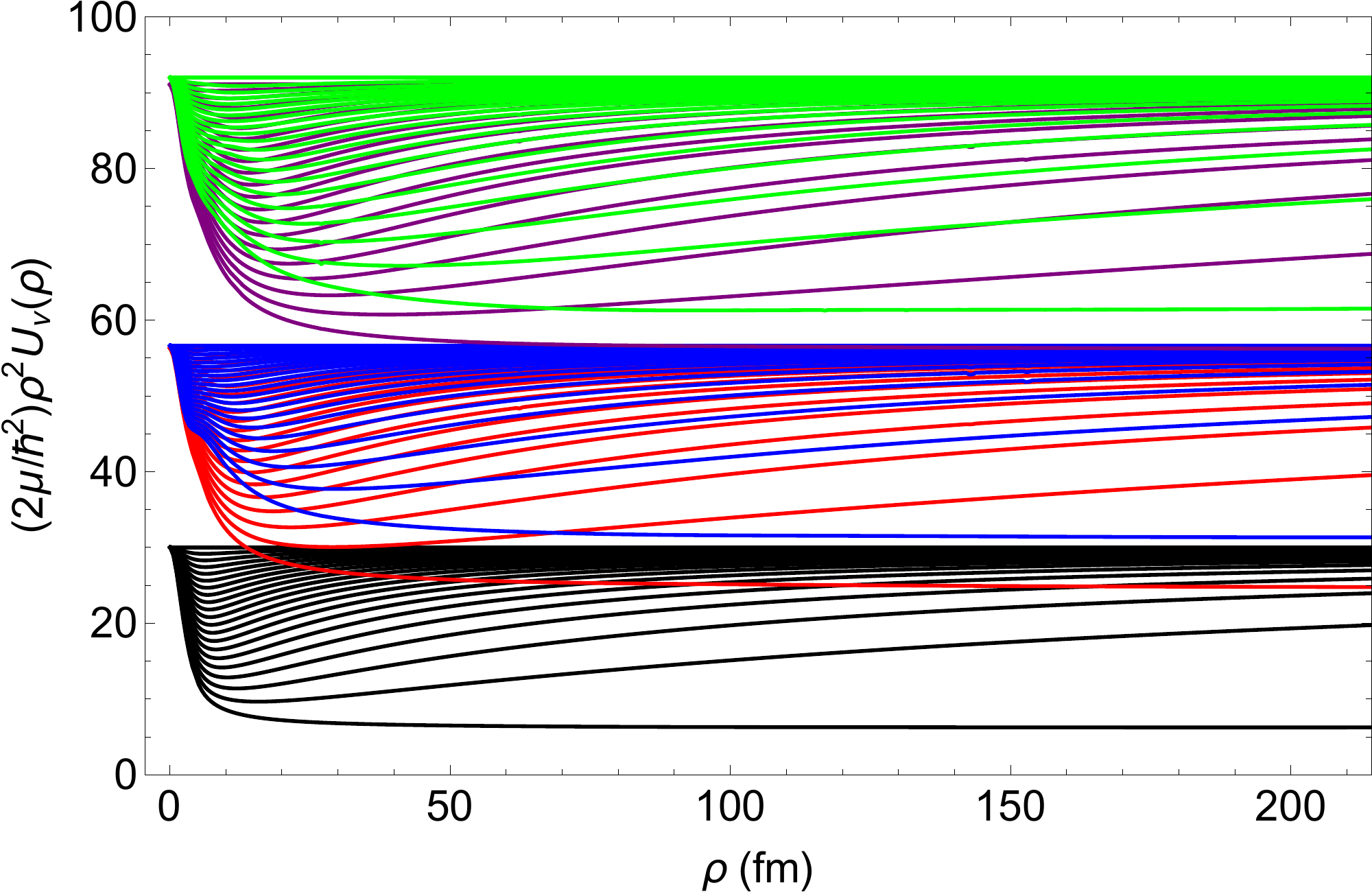}
    \caption{The lowest 5 adiabatic potentials for the 4n system for the $J^{\pi}=0^{+}$ symmetry with $L=0$ using a single--Gaussian potential with $r_{0}=1.78$~fm. The strength $V_0$ is tuned to give different scattering lengths from 0 to infinity with no two--body bound state. Each set of potentials represent a different $s$--wave scattering length in this range.}
    \label{fig:C3_Coeff4n}
\end{figure}

To study the long--range behavior of the potentials, a single Gaussian model for the two--body interaction is used of the form,
\begin{equation}
    V(r)=V_0\mathrm{exp}\biggr(-\frac{r^2}{r_0^2}\biggr).
\end{equation}
For a fixed range $r_0$, the strength $V_0$ is tuned to give a different $s$--wave scattering length, while maintaining the condition that there is no two--body bound state. The values of $V_0$ are chosen to give scattering lengths ranging from zero to infinity, coinciding with the non--interacting case to the unitarity regime. To represent the nn interaction, the values of $V_0$ and $r_0$ are chosen to reproduce the dominant low energy $s$--wave and $p$--wave properties produced by the spin--singlet and triplet components of the central term in the $\mathrm{AV8}^{\prime}$ interaction. The parameters used in the single--Gaussian calculations in Section \ref{sec:Adiabatic_Potentials} are given in Table \ref{table:gauss_params}, along with the strength associated with the unitarity limit for the spin--singlet state denoted by $V_{u}$. 
\begin{table}[ht]
\caption{Single Gaussian parameters used for singlet and triplet two--body interactions. The parameters were extracted from fits to the central component of the $\mathrm{AV8}^{\prime}$ potential. The parameter $V_u$ is the strength at unitarity.} 
\centering
\label{table:gauss_params}
\resizebox{0.8\columnwidth}{!}{\begin{tabular}{l|l|l|l}
    $\mathrm{State}$ & $V_{0} (\mathrm{MeV})$ & $V_{u} (\mathrm{MeV})$ & $r_{0} (\mathrm{fm})$\\
    \hline
    $\prescript{1}{}{S}$ & $-31.7674$ & $-35.1265$ & $1.7801$\\
    $\prescript{3}{}{P}$ & $95.7280$ & $-646.625$ &$0.8809$ \\\hline 
\end{tabular}}
\end{table}
To extract the expected value of $C_{3,\nu}$ for the few neutron systems, the strength of the spin--singlet state is tuned from $0<V_{0}<V_{u}$ with $r_0$ fixed to the value given in the first row of Table \ref{table:gauss_params}.
The lowest few adiabatic potentials for this study are shown in Figures \ref{fig:C3_Coeff3n} and \ref{fig:C3_Coeff4n} for the 3n and 4n systems, respectively.
These figures show the lowest few potentials for 20 scattering lengths ranging from 0 to unitarity for $r_0=1.7801$ fm, plotted as $(2\mu/\hbar^{2})\rho^{2}U(\rho)$ versus $\rho$. Each curve approaches the value $l_{\mathrm{eff}}(l_{\mathrm{eff}}+1)$ at large $\rho$ where, in the non--interacting limit, $l_{\mathrm{eff}}=5/2$ for the ground--state of the 3n $J^{\pi}=\frac{3}{2}^{-}$ system and $l_{\mathrm{eff}}=5$ for the ground--state of the 4n $J^{\pi}=0^{+}$ system. The lowest of each set of curves is the unitarity potential where the strength of the interaction is tuned to give an infinite scattering length. The unitarity values for $l_{\mathrm{eff}}$ describing the long--range potential are given in Table \ref{table:C3_coeff} and labeled as $l_{\mathrm{eff},\mathrm{u}}$. One general and universal property of every N-particle system with finite range interactions is that the long-range hyperradial potential curves associated with the N-body continuum can often (but not necessarily always) converge to a different asymptotic coefficient of $1/\rho^2$ at unitarity ($a\rightarrow \infty$) than for finite or vanishing scattering length.  This can be viewed as a generalized consequence of Efimov physics \cite{DIncaoReview,Rittenhouse_2011,Greene2017RMP,NaidonReview}. It is apparent from these figures that the lowest few potentials in each system asymptotically approach the corresponding non--interacting potentials as $\rho^{-3}$, where $a$ is finite.

\begin{figure}[h]
\centering
\subfigure[]{\includegraphics[width=8.5cm]{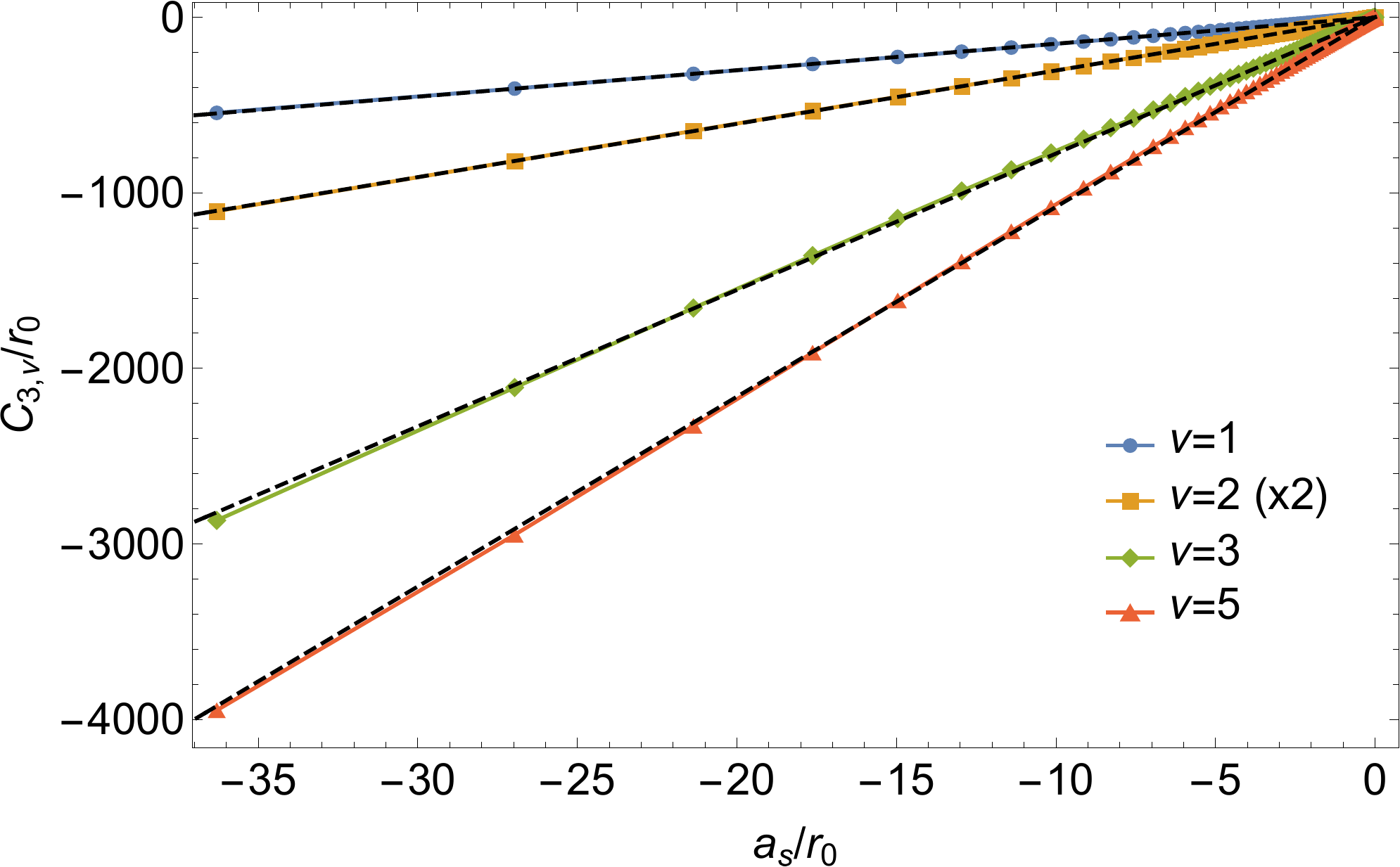}}

\subfigure[]{\includegraphics[width=8.5cm]{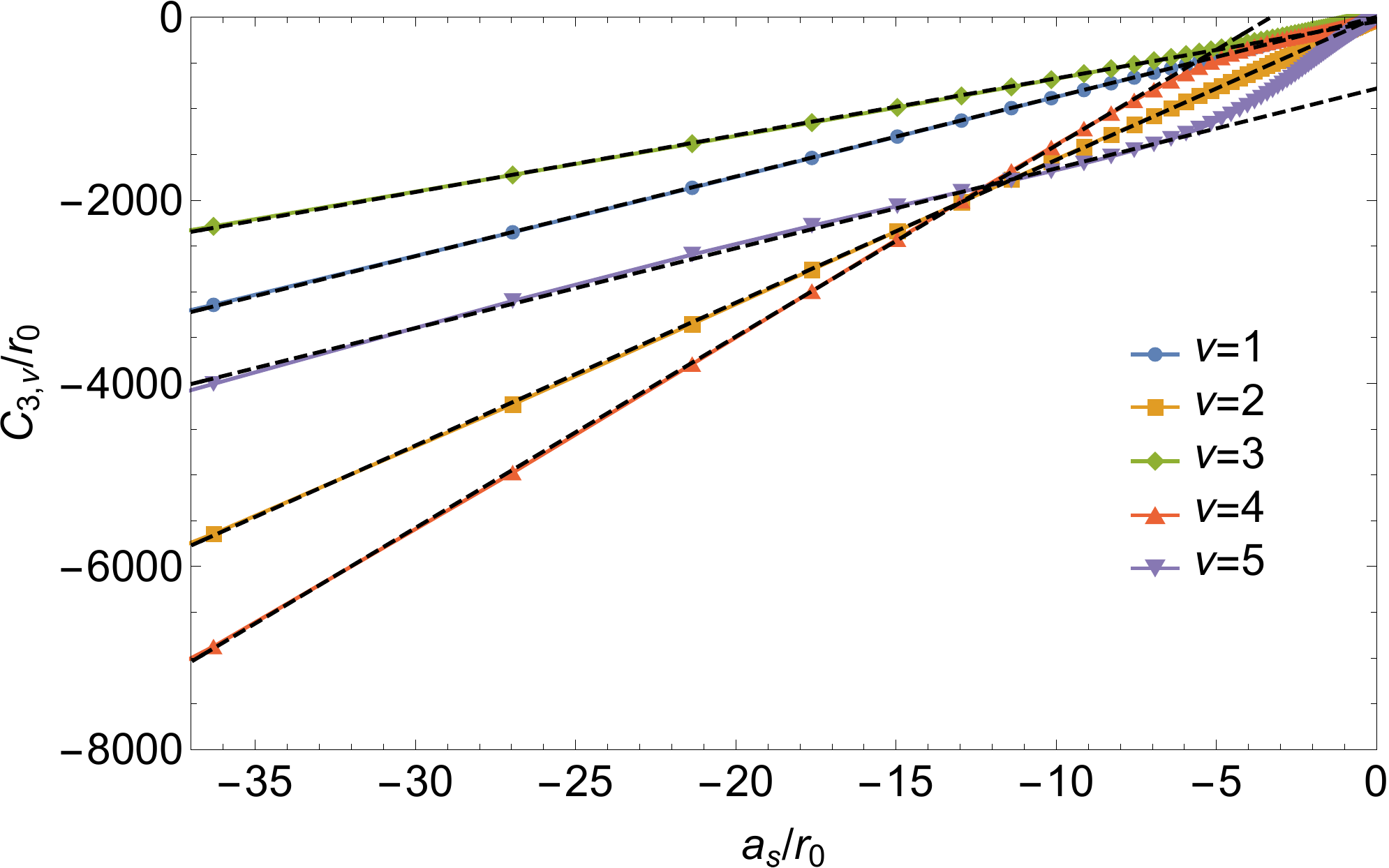}}
\caption{The scattering length dependence of the 3--body and 4--body $C_{3,\nu}$ coefficient scaled by the range of the potential, $r_0$ for the lowest few potentials. In (a), the re-scaled $C_{3,\nu}$ coefficient is shown for the 3--body case for the $(1\frac{1}{2})\frac{3}{2}^{-}$ symmetry. The circle, square, diamond and triangle symbols represent the channels $v=1,2,3, \mathrm{and}~5$, respectively. In (b), the re-scaled $C_{3,\nu}$ coefficient is shown for the 4--body case for the $(00)0^{+}$ symmetry. The circle, square, diamond, upright triangle, and downward triangle symbols represent the channels $v=1, 2, 3, 4, \mathrm{and}~5$, respectively. The dashed lines are linear fits performed over the range $-36<a_{s}/r_{0}<-10$.}
\label{fig:3Body_C3_Coeff}
\end{figure}

The scattering length dependence of $C_{3,\nu}$ is extracted from fitting the lowest adiabatic potentials to an inverse power--law expansion to $O(1/\rho^{8})$ for each scattering length from 10~fm~$<\rho<$~1500~fm. The results of $C_{3,\nu}$ for the 3n and 4n systems are shown in Figure \ref{fig:3Body_C3_Coeff}. 
The figures show the scattering length dependence of $C_{3,\nu}$, re-scaled by $r_0$, for the few lowest potentials using a single--Gaussian model for the two--body interactions with $r_{0}=1.78$~fm. There is clear evidence that the $C_{3,\nu}$ coefficient depends linearly on the $s$--wave scattering length for values of the re--scaled $s$--wave scattering length $|a_s/r_0|>10$, demonstrating universal physics at these scattering lengths. The re--scaled $s$--wave scattering length for the nn interaction is approximately, $|a_s/r_0|\approx10.63$. The numerical values of $C_{3,\nu}$ are given in Table \ref{table:C3_coeff}. The deviation from the linear fits (near $a_s/r_0$ of -30) are due to errors in the extraction on the $C_{3,\nu}$ coefficient, which were estimated by performing the fits over varying ranges of $|a_{s}/r_{0}|$. As the scattering length increases, the $\rho^{-3}$ behavior begins to dominate at larger and larger hyperradii, as seen in Figs. \ref{fig:C3_Coeff3n} and \ref{fig:C3_Coeff4n}. In order to improve the accuracy of the value of $C_{3,\nu}$, the potentials would need to be computed out to larger values of $\rho$ and better converged. Substituting the $s$--wave scattering length for the nn two--body interaction in the spin--singlet state ($a_{s}=-18.92$~fm for the $\mathrm{AV8}^{\prime}$ interaction) into the $C_{3,\nu}$ coefficients in Table \ref{table:C3_coeff} yields the expected long--range behavior for the lowest few adiabatic potentials that are relevant to understanding the low--energy behavior of the eigenphaseshifts, discussed in Sec. \ref{sec:wigner_smith}.

\begin{table}[ht]
\caption{Long--range ($\rho \rightarrow \infty$) $C_{\nu}$ coefficients of the lowest few adiabatic potential for the 3n and 4n systems (see Eq.\eqref{eq:asymform}). Also provided is the effective angular momentum $l_{\mathrm{eff},\mathrm{u}}$ at unitarity. Unitarity values from other references are presented as $l_{\mathrm{eff},\mathrm{u}}^{(\mathrm{ref})}$ for comparison. The error estimates in $C_{\nu}$ and $l_{\mathrm{eff},\mathrm{u}}$ are obtained from fitting the potentials over different ranges in the hyperradius.}
\centering
\label{table:C3_coeff}
\resizebox{0.8\columnwidth}{!}{\begin{tabular}{l|l|l|l|l|l|l}
    $N$ & $(LS)J^{\pi}$ & $\nu$ & $l_{\mathrm{eff}}$ & $C_{\nu}$ & $l_{\mathrm{eff},\mathrm{u}}$ & $l_{\mathrm{eff},\mathrm{u}}^{\mathrm{(ref)}}$\\ 
    \hline
    $3$ & $(1\frac{1}{2})\frac{3}{2}^-$ & $1$ &$5/2$ & $15.1(3)$ & $1.275(5)$ & $1.2727(1)\footnote{Values extracted from Table III of \cite{YinBlume2015pra}; The ground--state energies of two--component Fermi gases at unitarity are extracted in the zero--range limit.\label{fn:YinBlume}}, 1.2727^{\mathrm{\ref{fn:castin}}}$\\
    $~$ & $~$ & $2$ & $9/2$ & $15.2(3)$ & $3.861(5)$ & $3.868\footnote{Values extracted from Table I of \cite{vonStecher2007prl}; Energies of a trapped two--component Fermi gas are computed using hyperspherical techniques and given for a Gaussian interaction with a range of 0.05 oscillator units.\label{fn:stecher2007}}, 3.8582^{\mathrm{\ref{fn:castin}}}$\\
    $~$ & $~$ & $3$ & $13/2$ & $77.7(3)$ & $5.219(5)$ & $5.229^{\mathrm{\ref{fn:stecher2007}}}, 5.2164^{\mathrm{\ref{fn:castin}}}$\\
    $~$ & $~$ & $5$ & $17/2$ & $108(3)$ & $7.555(5)$ & $7.553\footnote{Value extracted from the transcendental equation represented by Eq. (7) in \cite{PhysRevLett.97.150401} for equal--mass fermions.\label{fn:castin}}$\\\hline
    $4$ & $(00)0^+$ & $1$ & $5$ & $86.7(3)$ & $2.017(5)$ & $2.0091(4)^{\mathrm{\ref{fn:YinBlume}}}$\\
    $~$ & $~$ & $2$ & $7$ & $156(3)$ & $4.455(5)$ & $4.444(3)\footnote{Values extracted from Table II of \cite{stecher2009PRA}; Long--range coefficients extracted from hyperspherical potential curves for a four--fermion system using a Gaussian two--body interaction at unitarity. \label{fn:stecher2009}}$\\
    $~$ & $~$ & $3$ & $7$ & $61.1(3)$ & $5.071(5)$ & $5.029(3)^{\mathrm{\ref{fn:stecher2009}}}$\\
    $~$ & $~$ & $4$ & $9$ & $209(3)$ & $6.974(5)$ & $6.863(3)^{\mathrm{\ref{fn:stecher2009}}}$\\
    $~$ & $~$ & $5$ & $9$ & $87.8(3)$ & $7.258(5)$ & $7.121(3)^{\mathrm{\ref{fn:stecher2009}}}$\\\hline
\end{tabular}}
\end{table}

\section{Wigner-Smith Time Delay}
\label{sec:wigner_smith}

Starting from this understanding of the long--range behavior of the adiabatic potentials demonstrated in Sec.~\ref{sec:potentials_long_range}, a full treatment of the energy--dependence scattering above the 3--body and 4--body continuum is now developed. To understand the 2016 experimental observation of a low--energy 4n signal by Kisamori et al., a detailed analysis of the low--energy density of states is carried out. A simple way of quantifying the density of states is through a calculation of the Wigner--Smith time delay matrix, which for a single--channel calculation represents the amount of time incoming probability flux remains confined by the presence of a potential before escaping \cite{Smith1960PR}. The Hermitian time delay matrix is defined in terms of the scattering matrix and its energy derivative as $Q(E)=i\hbar SdS^{\dagger}/dE$, which reduces to $Q(E)=2\hbar d\delta(E)/dE$ for a single--channel calculation. It has been shown there is a direct relation between the total Wigner--Smith time delay and the density of states \cite{TEXIER201616,OrangeReview}. A peak in the time delay results from a rapid increase in the phaseshift, usually over a small energy range. This occurs when the probability flux gets temporarily trapped for an extended period of time, interpreted usually as a resonance if there is an increase in phase of more than 2 radians.

\begin{figure}[!ht]
    \centering
    \includegraphics[width=8.5cm]{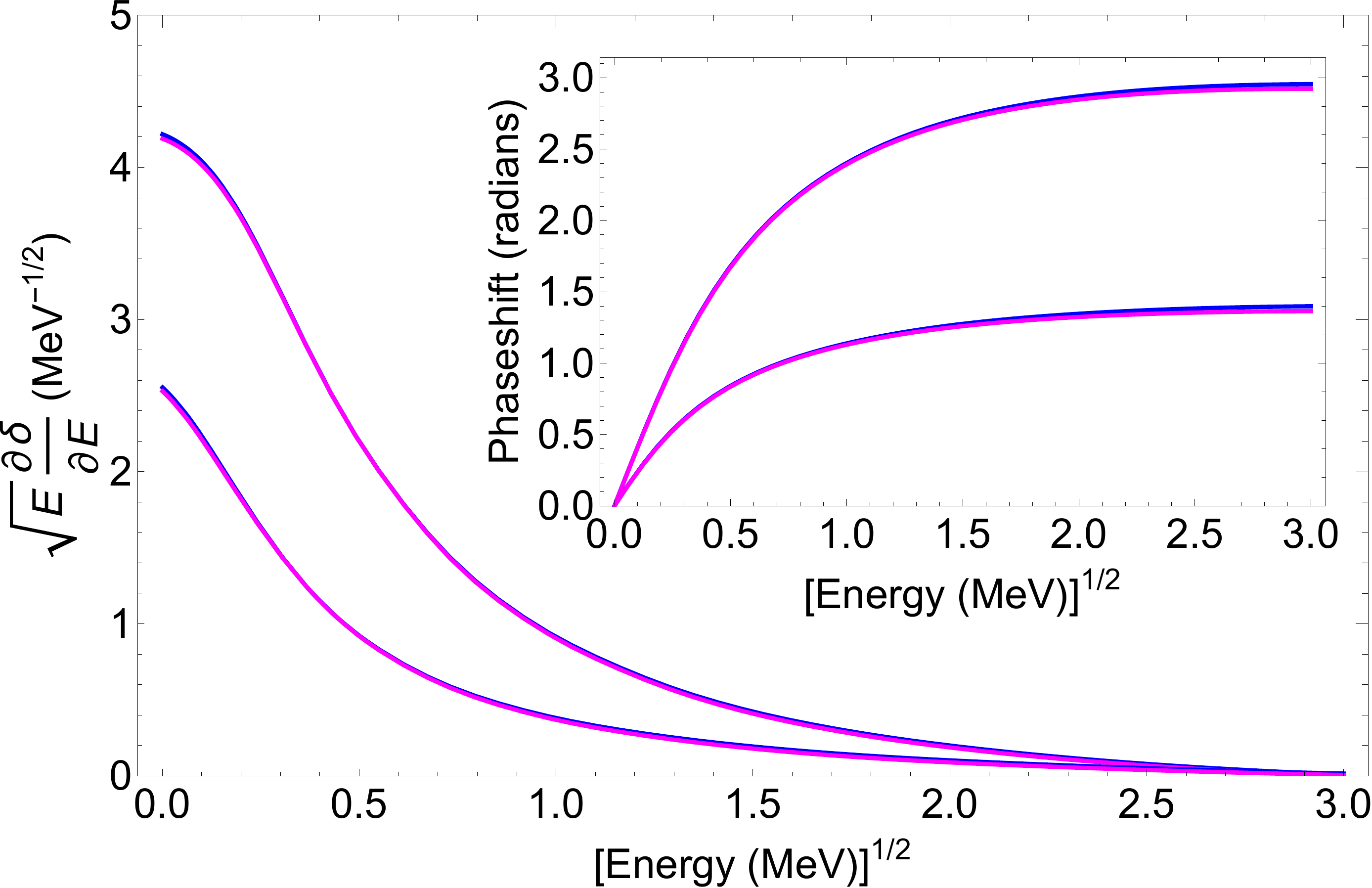}
    \caption{The elastic phaseshift(inset) and re--scaled Wigner--Smith time delay in the lowest adiabatic potential for the 3n and 4n systems with and without the diagonal second--derivative coupling term shown as magenta and blue lines, respectively. The upper curves are for the 4n system and the lower curves are for the 3n system.}
    \label{fig:3n_elastic_phase}
\end{figure}

In studying the elastic phaseshift of the 3n and 4n systems, a comparison can be made between the phaseshift with and without the inclusion of the second derivative coupling term. Since the lowest adiabatic potential with the inclusion of the non--adiabatic diagonal second derivative term provides a rigorous upper--limit to the actual lowest potential, it would be intuitive to study how important this term is to the low--energy behavior. Also, a direct comparison will shed some light on the accuracy of treating the problem in the adiabatic approximation, without including any non--adiabatic couplings. This comparison for both 3n and 4n systems is shown in Figure \ref{fig:3n_elastic_phase}.

The single--channel elastic phaseshift is shown for the 3n (lower curve) and 4n (upper curve) systems. These calculations were performed with and without the second derivative non--adiabatic coupling, being represented by the magenta and blue curves, respectively. There is surprising little change in the energy--dependence of the elastic phaseshift in the energy range shown from $0<E<9$~MeV, with a noticeable difference starting around 2 MeV. There is a smooth rise in the elastic phaseshift for both the 3n and 4n systems with no peak in the energy--derivative as one would expect for a true resonance. Instead, there is a low energy enhancement of the density of states due to the divergent $1/\sqrt{E}$ behavior. The enhancement by $1/\sqrt{E}$ is a result of the long--range $\rho^{-3}$ term in the potential. In accordance with the Wigner threshold law, the phase shift scales proportionally to the wave number $k$, leading to the enhancement in the energy--derivative. In fact, it can be shown from the Born approximation that a potential of the form $C/\rho^3$ leads to a phaseshift in the $l_{\mathrm{th}}$ partial wave of $\delta_{l}(k)\approx -kC/[2l(l+1)]$.

From Fig. \ref{fig:3n_elastic_phase}, there is little change in the energy behavior of the elastic phaseshift and time delay for these few--neutron systems when excluding the non--adiabatic couplings, treating the problem in a purely adiabatic sense. In fact, only at large scattering energies ($E>1.5$~MeV) is there any noticeable deviation in the energy dependence. The difference is due to the fact that the non--adiabatic second derivative couplings for both systems are only appreciable at small $\rho$, corresponding to large energy, as seen in the previous section. A more thorough investigation into the affects of the non--adiabatic couplings is done by performing a multi--channel calculation, including a few excited channels in solving Eq. \eqref{eq:coupled} for scattering solutions.

\begin{figure}[!ht]
    \centering
    \includegraphics[width=8.5cm]{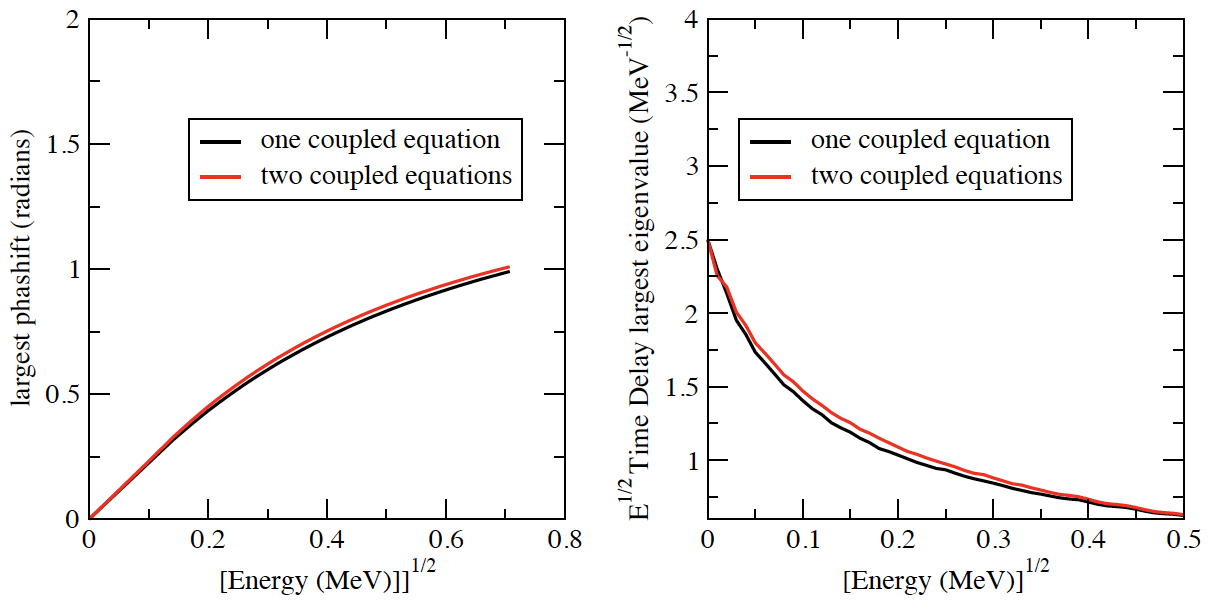}
    \caption{The largest 3n eigenphaseshift (left) and time delay (right) with the inclusion of 1 and 2 channels from a multi--channel calculation to show the effects on the low--energy behavior from non--adiabatic coupling. These results use a single Gaussian model for the nn two--body interaction with the HH basis.}
    \label{fig:multichannel_phase}
\end{figure}

\begin{figure}[!ht]
    \centering
    \includegraphics[width=8.5cm]{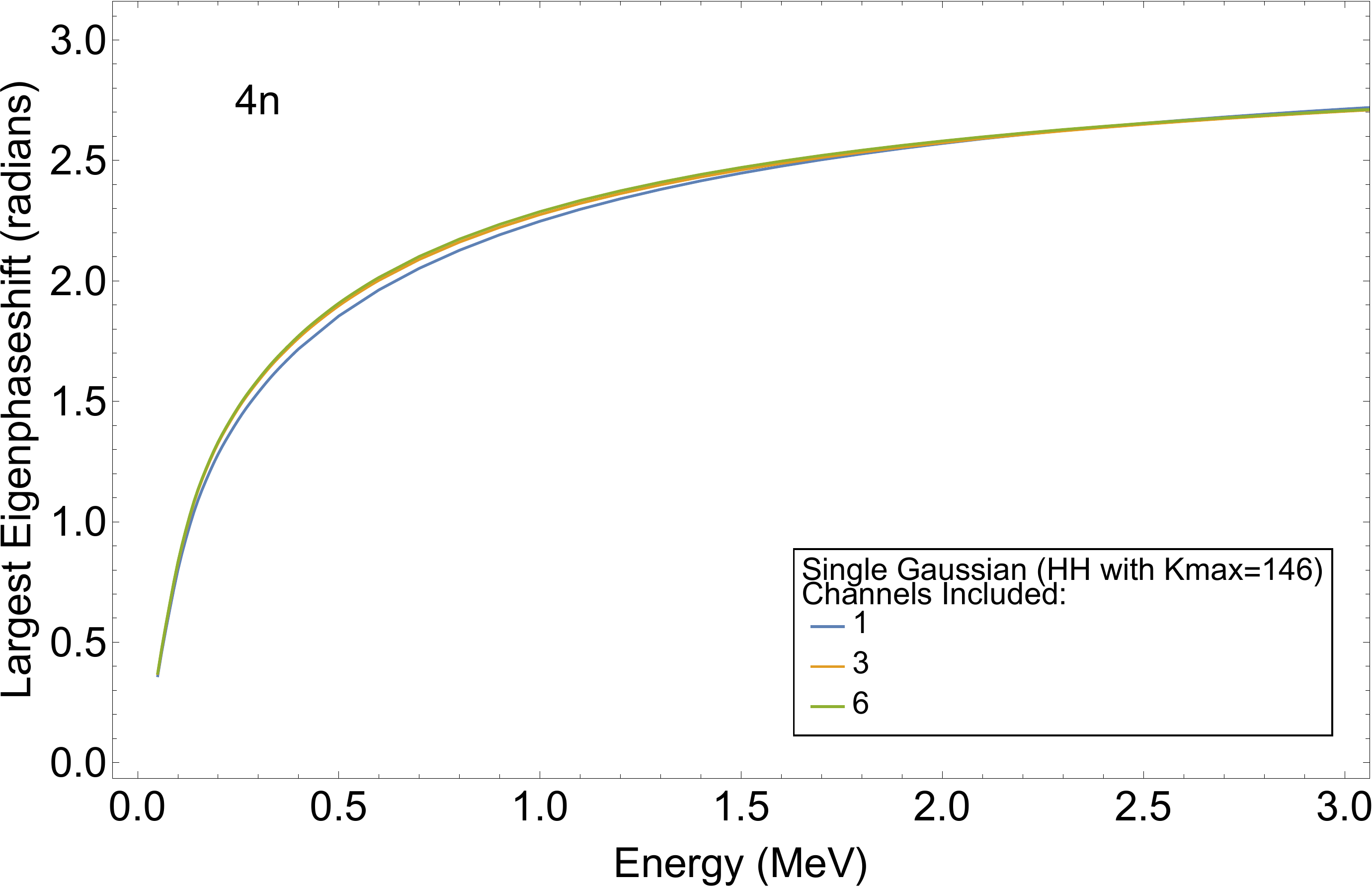}
    \caption{The 4n phaseshift in the lowest adiabatic potential with the inclusion of multiple channels an a multi--channel calculation to show the effects on the low--energy behavior from channel coupling. These results use a single Gaussian model for the nn two--body interaction with the HH basis.}
    \label{fig:multichannel_time}
\end{figure}

\begin{figure}[!ht]
    \centering

    \includegraphics[width=8.2 cm]{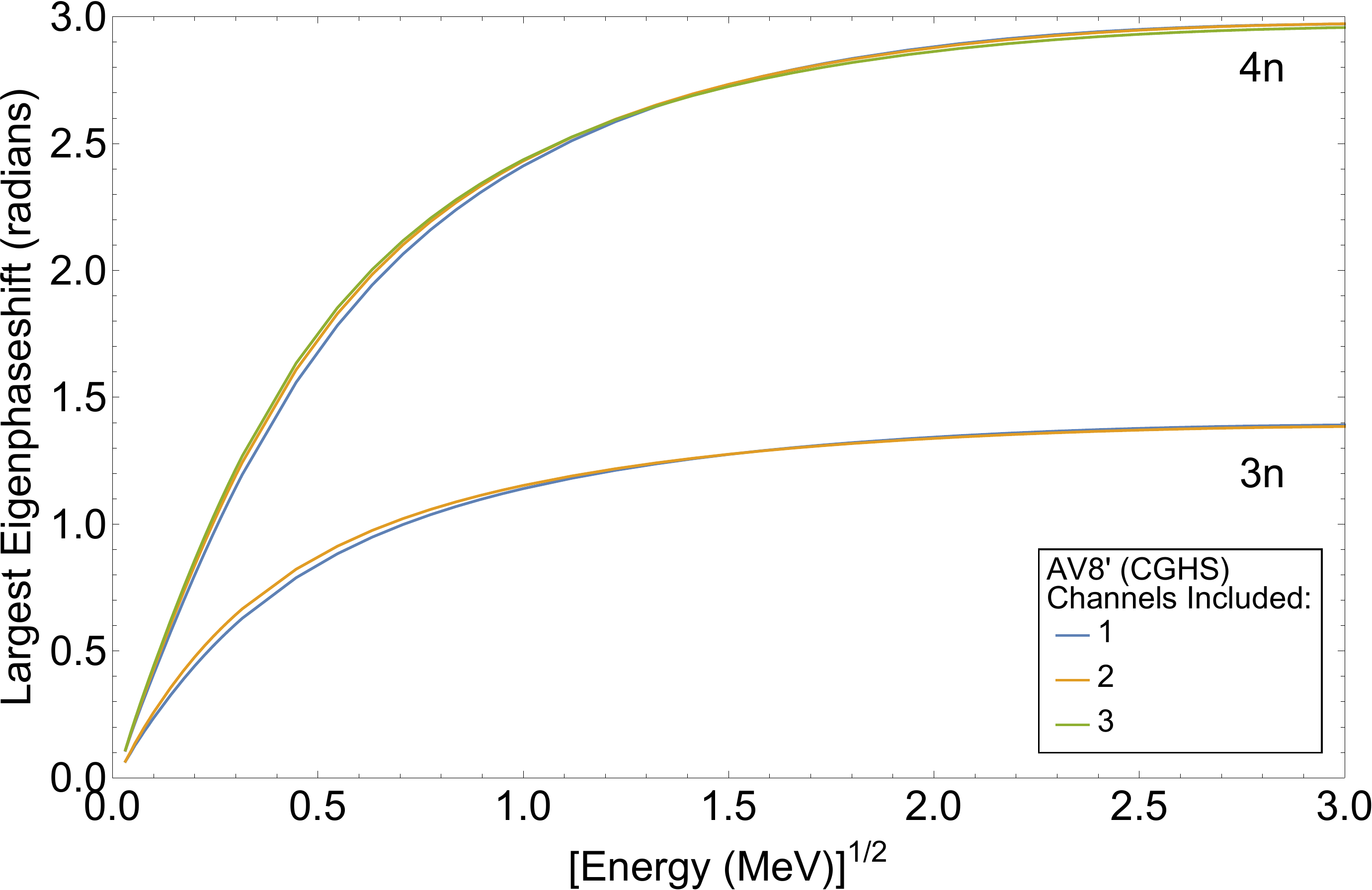}
    
    \caption{Largest eigenphaseshift for the 3n and 4n systems for the inclusion of up to 3 channels using the $\mathrm{AV8}^{\prime}$ interaction with the CGHS basis. The lowest curve is the single--channel results while the second lowest and highest curves are the resultant eigenphaseshifts from a mutichannel calculation for 2 and 3 channels, respectively. The eigenphaseshift is only slightly modified for energies less than 1 MeV, providing further support the adiabatic approximation is sufficient in describing the low--energy behavior.}
    \label{fig:multichannel_phase_CGHS}
\end{figure}

Multichannel scattering calculations were performed for both neutron systems. Equation \eqref{eq:coupled} was solved for the inclusion of a few coupled channels to compare the largest eigenphaseshifts, which are shown in Figs. \ref{fig:multichannel_phase} and \ref{fig:multichannel_time} for up to 2 and 6 channels for the 3n and 4n systems, respectively using the HH basis and a single Gaussian nuclear interaction. It is clear from these figures that the largest eigenphaseshift shows little to no change in energy dependence at low--energies with increased number of included channels. The largest eigenphaseshift using the $\mathrm{AV8}^{\prime}$ nuclear interaction and the CGHS basis is shown for up to 3 channels in Fig. \ref{fig:multichannel_phase_CGHS}. From these calculations with the realistic $\mathrm{AV8}^{\prime}$ interaction, the largest eigenphaseshift does not change significantly at low energies with the inclusion of more channels.  This negligible change is a result of the relatively weak and short--range behavior of the non--adiabatic coupling terms. From section \ref{sec:non_adiacatic_couplings}, it is shown the first--derivative couplings fall off as $\rho^{-2}$ and second--derivative couplings fall off as $\rho^{-4}$ at large hyperradii.

\section{Conclusions}
\label{sec:conclusions}
This article addresses a fundamental problem in few--body neutron interactions, whether a bound or resonant 4n state exists. The answer to this question is crucial for addressing the interpretation of a recent experiment by Kisamori et al. in 2016, which suggested evidence of a 4n resonance observed in the nuclear reaction $\prescript{4}{}{\mathrm{He}}+\prescript{8}{}{\mathrm{He}}\rightarrow4\mathrm{n}+\prescript{8}{}{\mathrm{Be}}$ at an energy of 1.25 MeV above the 4--neutron continuum \cite{Kisamori}. This problem has been treated here using the adiabatic hyperspherical approach with realistic NN and 3N interactions with an emphasis on the the $\mathrm{AV8}^{\prime}$ nuclear interaction with a CGHS basis and the AV18 interaction with a HH basis.

The 3n and 4n adiabatic hyperspherical potential curves were computed for the $J^{\pi}$ states that provide the most attraction in each system, $\frac{3}{2}^{-}$ for the 3n system and $0^{+}$ for the 4n system. A comparison was made between the adiabatic potentials computed in both basis sets, which show that the CGHS basis yields better converged potentials at large hyperradii. The lowest hyperspherical potential in each of these systems shows no qualitative feature that indicates the possible existence of a resonance. In fact, while these few neutron systems exhibit significant attraction, the potentials are still repulsive at all hyperradii due to strong Pauli repulsion.

The non-resonant behavior of these few neutron systems is further substantiated by a multi--channel scattering treatment in the 3--body and 4--body continua with an analysis of the eigenphaseshifts and Wigner--Smith time delay or density of states. It has been shown that the long--range behavior of the adiabatic potentials deviates from the non--interacting $\rho^{-2}$ by an attractive $\rho^{-3}$ behavior proportional to the $s$--wave scattering length. This attractive long--range term greatly influences the low--energy behavior of the eigenphaseshifts and in--turn, leading to a $1/\sqrt{E}$ enhancement of dependence of the density of states at low energy. This enhancement of the density of states could suggest an explanation for the low--energy signature observed in the Kisamori et al. experiment in 2016 \cite{Kisamori}, despite the fact that there is no peak in the density of states and no rapid increase in the eigenphaseshift of the type associated with a resonance. A multichannel treatment of few--neutron scattering was performed to include non--adiabatic coupling to excited states. This treatment showed little change in the energy--dependent phaseshift in the energy region of interest, which demonstrates that these systems can be accurately treated in a non--adiabatic picture.

Universal physics has also been studied and shown to be relevant for understanding these few neutrons systems due to the large size of the $s$--wave scattering length relative to the range of the nn interaction. The fact that the adiabatic potentials computed using a simple Gaussian interaction agree accurately with those using the full $\mathrm{AV8}^{\prime}$ potential provides strong evidence that these systems exhibit universal physics. Furthermore, the 3n and 4n systems have also been treated with different realistic nucleon--nucleon interactions from chiral effective field theory using the HH basis. The potentials using these interaction models all show qualitative agreement in the region of hyperradii where the HH basis is well converged. The universal behavior in these systems has been elucidated by studying the scattering length dependence of the long--range behavior of the lowest few adiabatic potentials using a Gaussian two--body interaction. At large ratios of the $s$--wave scattering length to the range of the potential ($|a_s/r_0|\ge10$), the long--range $\rho^{-3}$ coefficient depends linearly on the scattering length, consistent with what is expected with a delta--function contact potential.

\begin{acknowledgments}
Discussions with Emiko Hiyama and access to her unpublished Gaussian fitted $\mathrm{AV8}^{\prime}$ potential are much appreciated. The work done by MDH and CHG is supported in part by the U.S. National Science Foundation, Grant No.PHY-1912350, and in part by the Purdue Quantum Science and Engineering Institute.
\end{acknowledgments}

\bibliography{tetraneutron}

\providecommand{\noopsort}[1]{}\providecommand{\singleletter}[1]{#1}%
\begin{thebibliography}{57}%
\makeatletter
\providecommand \@ifxundefined [1]{%
 \@ifx{#1\undefined}
}%
\providecommand \@ifnum [1]{%
 \ifnum #1\expandafter \@firstoftwo
 \else \expandafter \@secondoftwo
 \fi
}%
\providecommand \@ifx [1]{%
 \ifx #1\expandafter \@firstoftwo
 \else \expandafter \@secondoftwo
 \fi
}%
\providecommand \natexlab [1]{#1}%
\providecommand \enquote  [1]{``#1''}%
\providecommand \bibnamefont  [1]{#1}%
\providecommand \bibfnamefont [1]{#1}%
\providecommand \citenamefont [1]{#1}%
\providecommand \href@noop [0]{\@secondoftwo}%
\providecommand \href [0]{\begingroup \@sanitize@url \@href}%
\providecommand \@href[1]{\@@startlink{#1}\@@href}%
\providecommand \@@href[1]{\endgroup#1\@@endlink}%
\providecommand \@sanitize@url [0]{\catcode `\\12\catcode `\$12\catcode
  `\&12\catcode `\#12\catcode `\^12\catcode `\_12\catcode `\%12\relax}%
\providecommand \@@startlink[1]{}%
\providecommand \@@endlink[0]{}%
\providecommand \url  [0]{\begingroup\@sanitize@url \@url }%
\providecommand \@url [1]{\endgroup\@href {#1}{\urlprefix }}%
\providecommand \urlprefix  [0]{URL }%
\providecommand \Eprint [0]{\href }%
\providecommand \doibase [0]{https://doi.org/}%
\providecommand \selectlanguage [0]{\@gobble}%
\providecommand \bibinfo  [0]{\@secondoftwo}%
\providecommand \bibfield  [0]{\@secondoftwo}%
\providecommand \translation [1]{[#1]}%
\providecommand \BibitemOpen [0]{}%
\providecommand \bibitemStop [0]{}%
\providecommand \bibitemNoStop [0]{.\EOS\space}%
\providecommand \EOS [0]{\spacefactor3000\relax}%
\providecommand \BibitemShut  [1]{\csname bibitem#1\endcsname}%
\let\auto@bib@innerbib\@empty
\bibitem [{\citenamefont {Higgins}\ \emph {et~al.}(2020)\citenamefont
  {Higgins}, \citenamefont {Greene}, \citenamefont {Kievsky},\ and\
  \citenamefont {Viviani}}]{PhysRevLett.125.052501}%
  \BibitemOpen
  \bibfield  {author} {\bibinfo {author} {\bibfnamefont {M.~D.}\ \bibnamefont
  {Higgins}}, \bibinfo {author} {\bibfnamefont {C.~H.}\ \bibnamefont {Greene}},
  \bibinfo {author} {\bibfnamefont {A.}~\bibnamefont {Kievsky}}, and\ \bibinfo
  {author} {\bibfnamefont {M.}~\bibnamefont {Viviani}},\ }\bibfield  {title}
  {\bibinfo {title} {Nonresonant density of states enhancement at low energies
  for three or four neutrons},\ }\href
  {https://doi.org/10.1103/PhysRevLett.125.052501} {\bibfield  {journal}
  {\bibinfo  {journal} {Phys. Rev. Lett.}\ }\textbf {\bibinfo {volume} {125}},\
  \bibinfo {pages} {052501} (\bibinfo {year} {2020})}\BibitemShut {NoStop}%
\bibitem [{\citenamefont {Kisamori}\ \emph {et~al.}(2016)\citenamefont
  {Kisamori}, \citenamefont {Shimoura}, \citenamefont {Miya}, \citenamefont
  {Michimasa}, \citenamefont {Ota}, \citenamefont {Assie}, \citenamefont
  {Baba}, \citenamefont {Baba}, \citenamefont {Beaumel}, \citenamefont
  {Dozono}, \citenamefont {Fujii}, \citenamefont {Fukuda}, \citenamefont {Go},
  \citenamefont {Hammache}, \citenamefont {Ideguchi}, \citenamefont {Inabe},
  \citenamefont {Itoh}, \citenamefont {Kameda}, \citenamefont {Kawase},
  \citenamefont {Kawabata}, \citenamefont {Kobayashi}, \citenamefont {Kondo},
  \citenamefont {Kubo}, \citenamefont {Kubota}, \citenamefont
  {Kurata-Nishimura}, \citenamefont {Lee}, \citenamefont {Maeda}, \citenamefont
  {Matsubara}, \citenamefont {Miki}, \citenamefont {Nishi}, \citenamefont
  {Noji}, \citenamefont {Sakaguchi}, \citenamefont {Sakai}, \citenamefont
  {Sasamoto}, \citenamefont {Sasano}, \citenamefont {Sato}, \citenamefont
  {Shimizu}, \citenamefont {Stolz}, \citenamefont {Suzuki}, \citenamefont
  {Takaki}, \citenamefont {Takeda}, \citenamefont {Takeuchi}, \citenamefont
  {Tamii}, \citenamefont {Tang}, \citenamefont {Tokieda}, \citenamefont
  {Tsumura}, \citenamefont {Uesaka}, \citenamefont {Yako}, \citenamefont
  {Yanagisawa}, \citenamefont {Yokoyama},\ and\ \citenamefont
  {Yoshida}}]{Kisamori}%
  \BibitemOpen
  \bibfield  {author} {\bibinfo {author} {\bibfnamefont {K.}~\bibnamefont
  {Kisamori}}, \bibinfo {author} {\bibfnamefont {S.}~\bibnamefont {Shimoura}},
  \bibinfo {author} {\bibfnamefont {H.}~\bibnamefont {Miya}}, \bibinfo {author}
  {\bibfnamefont {S.}~\bibnamefont {Michimasa}}, \bibinfo {author}
  {\bibfnamefont {S.}~\bibnamefont {Ota}}, \bibinfo {author} {\bibfnamefont
  {M.}~\bibnamefont {Assie}}, \bibinfo {author} {\bibfnamefont
  {H.}~\bibnamefont {Baba}}, \bibinfo {author} {\bibfnamefont {T.}~\bibnamefont
  {Baba}}, \bibinfo {author} {\bibfnamefont {D.}~\bibnamefont {Beaumel}},
  \bibinfo {author} {\bibfnamefont {M.}~\bibnamefont {Dozono}}, \bibinfo
  {author} {\bibfnamefont {T.}~\bibnamefont {Fujii}}, \bibinfo {author}
  {\bibfnamefont {N.}~\bibnamefont {Fukuda}}, \bibinfo {author} {\bibfnamefont
  {S.}~\bibnamefont {Go}}, \bibinfo {author} {\bibfnamefont {F.}~\bibnamefont
  {Hammache}}, \bibinfo {author} {\bibfnamefont {E.}~\bibnamefont {Ideguchi}},
  \bibinfo {author} {\bibfnamefont {N.}~\bibnamefont {Inabe}}, \bibinfo
  {author} {\bibfnamefont {M.}~\bibnamefont {Itoh}}, \bibinfo {author}
  {\bibfnamefont {D.}~\bibnamefont {Kameda}}, \bibinfo {author} {\bibfnamefont
  {S.}~\bibnamefont {Kawase}}, \bibinfo {author} {\bibfnamefont
  {T.}~\bibnamefont {Kawabata}}, \bibinfo {author} {\bibfnamefont
  {M.}~\bibnamefont {Kobayashi}}, \bibinfo {author} {\bibfnamefont
  {Y.}~\bibnamefont {Kondo}}, \bibinfo {author} {\bibfnamefont
  {T.}~\bibnamefont {Kubo}}, \bibinfo {author} {\bibfnamefont {Y.}~\bibnamefont
  {Kubota}}, \bibinfo {author} {\bibfnamefont {M.}~\bibnamefont
  {Kurata-Nishimura}}, \bibinfo {author} {\bibfnamefont {C.~S.}\ \bibnamefont
  {Lee}}, \bibinfo {author} {\bibfnamefont {Y.}~\bibnamefont {Maeda}}, \bibinfo
  {author} {\bibfnamefont {H.}~\bibnamefont {Matsubara}}, \bibinfo {author}
  {\bibfnamefont {K.}~\bibnamefont {Miki}}, \bibinfo {author} {\bibfnamefont
  {T.}~\bibnamefont {Nishi}}, \bibinfo {author} {\bibfnamefont
  {S.}~\bibnamefont {Noji}}, \bibinfo {author} {\bibfnamefont {S.}~\bibnamefont
  {Sakaguchi}}, \bibinfo {author} {\bibfnamefont {H.}~\bibnamefont {Sakai}},
  \bibinfo {author} {\bibfnamefont {Y.}~\bibnamefont {Sasamoto}}, \bibinfo
  {author} {\bibfnamefont {M.}~\bibnamefont {Sasano}}, \bibinfo {author}
  {\bibfnamefont {H.}~\bibnamefont {Sato}}, \bibinfo {author} {\bibfnamefont
  {Y.}~\bibnamefont {Shimizu}}, \bibinfo {author} {\bibfnamefont
  {A.}~\bibnamefont {Stolz}}, \bibinfo {author} {\bibfnamefont
  {H.}~\bibnamefont {Suzuki}}, \bibinfo {author} {\bibfnamefont
  {M.}~\bibnamefont {Takaki}}, \bibinfo {author} {\bibfnamefont
  {H.}~\bibnamefont {Takeda}}, \bibinfo {author} {\bibfnamefont
  {S.}~\bibnamefont {Takeuchi}}, \bibinfo {author} {\bibfnamefont
  {A.}~\bibnamefont {Tamii}}, \bibinfo {author} {\bibfnamefont
  {L.}~\bibnamefont {Tang}}, \bibinfo {author} {\bibfnamefont {H.}~\bibnamefont
  {Tokieda}}, \bibinfo {author} {\bibfnamefont {M.}~\bibnamefont {Tsumura}},
  \bibinfo {author} {\bibfnamefont {T.}~\bibnamefont {Uesaka}}, \bibinfo
  {author} {\bibfnamefont {K.}~\bibnamefont {Yako}}, \bibinfo {author}
  {\bibfnamefont {Y.}~\bibnamefont {Yanagisawa}}, \bibinfo {author}
  {\bibfnamefont {R.}~\bibnamefont {Yokoyama}}, and\ \bibinfo {author}
  {\bibfnamefont {K.}~\bibnamefont {Yoshida}},\ }\bibfield  {title} {\bibinfo
  {title} {{Candidate Resonant Tetraneutron State Populated by the He-4 (He-8,
  Be-8) Reaction}},\ }\href {https://doi.org/{10.1103/PhysRevLett.116.052501}}
  {\bibfield  {journal} {\bibinfo  {journal} {{Phys. Rev. Lett.}}\ }\textbf
  {\bibinfo {volume} {{116}}},\ \bibinfo {pages} {{052501}} (\bibinfo {year}
  {{2016}})}\BibitemShut {NoStop}%
\bibitem [{\citenamefont {Marqu\'es}\ \emph {et~al.}(2002)\citenamefont
  {Marqu\'es}, \citenamefont {Labiche}, \citenamefont {Orr}, \citenamefont
  {Ang\'elique}, \citenamefont {Axelsson}, \citenamefont {Benoit},
  \citenamefont {Bergmann}, \citenamefont {Borge}, \citenamefont {Catford},
  \citenamefont {Chappell}, \citenamefont {Clarke}, \citenamefont {Costa},
  \citenamefont {Curtis}, \citenamefont {D'Arrigo}, \citenamefont
  {de~G\'oes~Brennand}, \citenamefont {de~Oliveira~Santos}, \citenamefont
  {Dorvaux}, \citenamefont {Fazio}, \citenamefont {Freer}, \citenamefont
  {Fulton}, \citenamefont {Giardina}, \citenamefont {Gr\'evy}, \citenamefont
  {Guillemaud-Mueller}, \citenamefont {Hanappe}, \citenamefont {Heusch},
  \citenamefont {Jonson}, \citenamefont {Le~Brun}, \citenamefont {Leenhardt},
  \citenamefont {Lewitowicz}, \citenamefont {L\'opez}, \citenamefont
  {Markenroth}, \citenamefont {Mueller}, \citenamefont {Nilsson}, \citenamefont
  {Ninane}, \citenamefont {Nyman}, \citenamefont {Piqueras}, \citenamefont
  {Riisager}, \citenamefont {Laurent}, \citenamefont {Sarazin}, \citenamefont
  {Singer}, \citenamefont {Sorlin},\ and\ \citenamefont
  {Stuttg\'e}}]{PhysRevC.65.044006}%
  \BibitemOpen
  \bibfield  {author} {\bibinfo {author} {\bibfnamefont {F.~M.}\ \bibnamefont
  {Marqu\'es}}, \bibinfo {author} {\bibfnamefont {M.}~\bibnamefont {Labiche}},
  \bibinfo {author} {\bibfnamefont {N.~A.}\ \bibnamefont {Orr}}, \bibinfo
  {author} {\bibfnamefont {J.~C.}\ \bibnamefont {Ang\'elique}}, \bibinfo
  {author} {\bibfnamefont {L.}~\bibnamefont {Axelsson}}, \bibinfo {author}
  {\bibfnamefont {B.}~\bibnamefont {Benoit}}, \bibinfo {author} {\bibfnamefont
  {U.~C.}\ \bibnamefont {Bergmann}}, \bibinfo {author} {\bibfnamefont
  {M.~J.~G.}\ \bibnamefont {Borge}}, \bibinfo {author} {\bibfnamefont {W.~N.}\
  \bibnamefont {Catford}}, \bibinfo {author} {\bibfnamefont {S.~P.~G.}\
  \bibnamefont {Chappell}}, \bibinfo {author} {\bibfnamefont {N.~M.}\
  \bibnamefont {Clarke}}, \bibinfo {author} {\bibfnamefont {G.}~\bibnamefont
  {Costa}}, \bibinfo {author} {\bibfnamefont {N.}~\bibnamefont {Curtis}},
  \bibinfo {author} {\bibfnamefont {A.}~\bibnamefont {D'Arrigo}}, \bibinfo
  {author} {\bibfnamefont {E.}~\bibnamefont {de~G\'oes~Brennand}}, \bibinfo
  {author} {\bibfnamefont {F.}~\bibnamefont {de~Oliveira~Santos}}, \bibinfo
  {author} {\bibfnamefont {O.}~\bibnamefont {Dorvaux}}, \bibinfo {author}
  {\bibfnamefont {G.}~\bibnamefont {Fazio}}, \bibinfo {author} {\bibfnamefont
  {M.}~\bibnamefont {Freer}}, \bibinfo {author} {\bibfnamefont {B.~R.}\
  \bibnamefont {Fulton}}, \bibinfo {author} {\bibfnamefont {G.}~\bibnamefont
  {Giardina}}, \bibinfo {author} {\bibfnamefont {S.}~\bibnamefont {Gr\'evy}},
  \bibinfo {author} {\bibfnamefont {D.}~\bibnamefont {Guillemaud-Mueller}},
  \bibinfo {author} {\bibfnamefont {F.}~\bibnamefont {Hanappe}}, \bibinfo
  {author} {\bibfnamefont {B.}~\bibnamefont {Heusch}}, \bibinfo {author}
  {\bibfnamefont {B.}~\bibnamefont {Jonson}}, \bibinfo {author} {\bibfnamefont
  {C.}~\bibnamefont {Le~Brun}}, \bibinfo {author} {\bibfnamefont
  {S.}~\bibnamefont {Leenhardt}}, \bibinfo {author} {\bibfnamefont
  {M.}~\bibnamefont {Lewitowicz}}, \bibinfo {author} {\bibfnamefont {M.~J.}\
  \bibnamefont {L\'opez}}, \bibinfo {author} {\bibfnamefont {K.}~\bibnamefont
  {Markenroth}}, \bibinfo {author} {\bibfnamefont {A.~C.}\ \bibnamefont
  {Mueller}}, \bibinfo {author} {\bibfnamefont {T.}~\bibnamefont {Nilsson}},
  \bibinfo {author} {\bibfnamefont {A.}~\bibnamefont {Ninane}}, \bibinfo
  {author} {\bibfnamefont {G.}~\bibnamefont {Nyman}}, \bibinfo {author}
  {\bibfnamefont {I.}~\bibnamefont {Piqueras}}, \bibinfo {author}
  {\bibfnamefont {K.}~\bibnamefont {Riisager}}, \bibinfo {author}
  {\bibfnamefont {M.~G.~S.}\ \bibnamefont {Laurent}}, \bibinfo {author}
  {\bibfnamefont {F.}~\bibnamefont {Sarazin}}, \bibinfo {author} {\bibfnamefont
  {S.~M.}\ \bibnamefont {Singer}}, \bibinfo {author} {\bibfnamefont
  {O.}~\bibnamefont {Sorlin}}, and\ \bibinfo {author} {\bibfnamefont
  {L.}~\bibnamefont {Stuttg\'e}},\ }\bibfield  {title} {\bibinfo {title}
  {Detection of neutron clusters},\ }\href
  {https://doi.org/10.1103/PhysRevC.65.044006} {\bibfield  {journal} {\bibinfo
  {journal} {Phys. Rev. C}\ }\textbf {\bibinfo {volume} {65}},\ \bibinfo
  {pages} {044006} (\bibinfo {year} {2002})}\BibitemShut {NoStop}%
\bibitem [{\citenamefont {Timofeyuk}(2003)}]{Timofeyuk_2003}%
  \BibitemOpen
  \bibfield  {author} {\bibinfo {author} {\bibfnamefont {N.~K.}\ \bibnamefont
  {Timofeyuk}},\ }\bibfield  {title} {\bibinfo {title} {Do multineutrons
  exist?},\ }\href {https://doi.org/10.1088/0954-3899/29/2/102} {\bibfield
  {journal} {\bibinfo  {journal} {Journal of Physics G: Nuclear and Particle
  Physics}\ }\textbf {\bibinfo {volume} {29}},\ \bibinfo {pages} {L9} (\bibinfo
  {year} {2003})}\BibitemShut {NoStop}%
\bibitem [{\citenamefont {Pieper}(2003)}]{PhysRevLett.90.252501}%
  \BibitemOpen
  \bibfield  {author} {\bibinfo {author} {\bibfnamefont {S.~C.}\ \bibnamefont
  {Pieper}},\ }\bibfield  {title} {\bibinfo {title} {Can modern nuclear
  hamiltonians tolerate a bound tetraneutron?},\ }\href
  {https://doi.org/10.1103/PhysRevLett.90.252501} {\bibfield  {journal}
  {\bibinfo  {journal} {Phys. Rev. Lett.}\ }\textbf {\bibinfo {volume} {90}},\
  \bibinfo {pages} {252501} (\bibinfo {year} {2003})}\BibitemShut {NoStop}%
\bibitem [{\citenamefont {Bertulani\text{ \ }}\ and\ \citenamefont
  {Zelevinsky}(2003)}]{Bertulani_2003}%
  \BibitemOpen
  \bibfield  {author} {\bibinfo {author} {\bibfnamefont {C.~A.}\ \bibnamefont
  {Bertulani\text{ \ }}}and\ \bibinfo {author} {\bibfnamefont {V.}~\bibnamefont
  {Zelevinsky}},\ }\bibfield  {title} {\bibinfo {title} {Is the tetraneutron a
  bound dineutron{\textendash}dineutron molecule?},\ }\href
  {https://doi.org/10.1088/0954-3899/29/10/309} {\bibfield  {journal} {\bibinfo
   {journal} {Journal of Physics G: Nuclear and Particle Physics}\ }\textbf
  {\bibinfo {volume} {29}},\ \bibinfo {pages} {2431} (\bibinfo {year}
  {2003})}\BibitemShut {NoStop}%
\bibitem [{\citenamefont {Shirokov}\ \emph {et~al.}(2016)\citenamefont
  {Shirokov}, \citenamefont {Papadimitriou}, \citenamefont {Mazur},
  \citenamefont {Mazur}, \citenamefont {Roth},\ and\ \citenamefont
  {Vary}}]{ShirokovVary2016prl}%
  \BibitemOpen
  \bibfield  {author} {\bibinfo {author} {\bibfnamefont {A.~M.}\ \bibnamefont
  {Shirokov}}, \bibinfo {author} {\bibfnamefont {G.}~\bibnamefont
  {Papadimitriou}}, \bibinfo {author} {\bibfnamefont {A.~I.}\ \bibnamefont
  {Mazur}}, \bibinfo {author} {\bibfnamefont {I.~A.}\ \bibnamefont {Mazur}},
  \bibinfo {author} {\bibfnamefont {R.}~\bibnamefont {Roth}}, and\ \bibinfo
  {author} {\bibfnamefont {J.~P.}\ \bibnamefont {Vary}},\ }\bibfield  {title}
  {\bibinfo {title} {{Prediction for a Four-Neutron Resonance}},\ }\href
  {https://doi.org/{10.1103/PhysRevLett.117.182502}} {\bibfield  {journal}
  {\bibinfo  {journal} {{Phys. Rev. Lett.}}\ }\textbf {\bibinfo {volume}
  {{117}}},\ \bibinfo {pages} {182502} (\bibinfo {year} {{2016}})}\BibitemShut
  {NoStop}%
\bibitem [{\citenamefont {Gandolfi}\ \emph {et~al.}(2017)\citenamefont
  {Gandolfi}, \citenamefont {Hammer}, \citenamefont {Klos}, \citenamefont
  {Lynn},\ and\ \citenamefont {Schwenk}}]{Gandolfi2017prc}%
  \BibitemOpen
  \bibfield  {author} {\bibinfo {author} {\bibfnamefont {S.}~\bibnamefont
  {Gandolfi}}, \bibinfo {author} {\bibfnamefont {H.-W.}\ \bibnamefont
  {Hammer}}, \bibinfo {author} {\bibfnamefont {P.}~\bibnamefont {Klos}},
  \bibinfo {author} {\bibfnamefont {J.~E.}\ \bibnamefont {Lynn}}, and\ \bibinfo
  {author} {\bibfnamefont {A.}~\bibnamefont {Schwenk}},\ }\bibfield  {title}
  {\bibinfo {title} {Is a trineutron resonance lower in energy than a
  tetraneutron resonance?},\ }\href
  {https://doi.org/10.1103/PhysRevLett.118.232501} {\bibfield  {journal}
  {\bibinfo  {journal} {Phys. Rev. Lett.}\ }\textbf {\bibinfo {volume} {118}},\
  \bibinfo {pages} {232501} (\bibinfo {year} {2017})}\BibitemShut {NoStop}%
\bibitem [{\citenamefont {Li}\ \emph {et~al.}(2019)\citenamefont {Li},
  \citenamefont {Michel}, \citenamefont {Hu}, \citenamefont {Zuo},\ and\
  \citenamefont {Xu}}]{LiMichelHu2019PRC}%
  \BibitemOpen
  \bibfield  {author} {\bibinfo {author} {\bibfnamefont {J.~G.}\ \bibnamefont
  {Li}}, \bibinfo {author} {\bibfnamefont {N.}~\bibnamefont {Michel}}, \bibinfo
  {author} {\bibfnamefont {B.~S.}\ \bibnamefont {Hu}}, \bibinfo {author}
  {\bibfnamefont {W.}~\bibnamefont {Zuo}}, and\ \bibinfo {author}
  {\bibfnamefont {F.~R.}\ \bibnamefont {Xu}},\ }\bibfield  {title} {\bibinfo
  {title} {{Ab initio no-core Gamow shell-model calculations of multineutron
  systems}},\ }\href {https://doi.org/{10.1103/PhysRevC.100.054313}} {\bibfield
   {journal} {\bibinfo  {journal} {{Phys. Rev. C}}\ }\textbf {\bibinfo {volume}
  {{100}}},\ \bibinfo {pages} {054313} (\bibinfo {year} {{2019}})}\BibitemShut
  {NoStop}%
\bibitem [{\citenamefont {Hiyama}\ \emph {et~al.}(2016)\citenamefont {Hiyama},
  \citenamefont {Lazauskas}, \citenamefont {Carbonell},\ and\ \citenamefont
  {Kamimura}}]{HiyamaLazauskas2016PRC}%
  \BibitemOpen
  \bibfield  {author} {\bibinfo {author} {\bibfnamefont {E.}~\bibnamefont
  {Hiyama}}, \bibinfo {author} {\bibfnamefont {R.}~\bibnamefont {Lazauskas}},
  \bibinfo {author} {\bibfnamefont {J.}~\bibnamefont {Carbonell}}, and\
  \bibinfo {author} {\bibfnamefont {M.}~\bibnamefont {Kamimura}},\ }\bibfield
  {title} {\bibinfo {title} {Possibility of generating a 4-neutron resonance
  with a $t=3/2$ isospin 3-neutron force},\ }\href
  {https://doi.org/10.1103/PhysRevC.93.044004} {\bibfield  {journal} {\bibinfo
  {journal} {Phys. Rev. C}\ }\textbf {\bibinfo {volume} {93}},\ \bibinfo
  {pages} {044004} (\bibinfo {year} {2016})}\BibitemShut {NoStop}%
\bibitem [{\citenamefont {Fossez}\ \emph {et~al.}(2017)\citenamefont {Fossez},
  \citenamefont {Rotureau}, \citenamefont {Michel},\ and\ \citenamefont
  {P\l{}oszajczak}}]{Fossez}%
  \BibitemOpen
  \bibfield  {author} {\bibinfo {author} {\bibfnamefont {K.}~\bibnamefont
  {Fossez}}, \bibinfo {author} {\bibfnamefont {J.}~\bibnamefont {Rotureau}},
  \bibinfo {author} {\bibfnamefont {N.}~\bibnamefont {Michel}}, and\ \bibinfo
  {author} {\bibfnamefont {M.}~\bibnamefont {P\l{}oszajczak}},\ }\bibfield
  {title} {\bibinfo {title} {Can tetraneutron be a narrow resonance?},\ }\href
  {https://doi.org/10.1103/PhysRevLett.119.032501} {\bibfield  {journal}
  {\bibinfo  {journal} {Phys. Rev. Lett.}\ }\textbf {\bibinfo {volume} {119}},\
  \bibinfo {pages} {032501} (\bibinfo {year} {2017})}\BibitemShut {NoStop}%
\bibitem [{\citenamefont {Deltuva}(2018)}]{Deltuva4n2018PL}%
  \BibitemOpen
  \bibfield  {author} {\bibinfo {author} {\bibfnamefont {A.}~\bibnamefont
  {Deltuva}},\ }\bibfield  {title} {\bibinfo {title} {Tetraneutron: Rigorous
  continuum calculation},\ }\href
  {https://doi.org/https://doi.org/10.1016/j.physletb.2018.05.041} {\bibfield
  {journal} {\bibinfo  {journal} {Physics Letters B}\ }\textbf {\bibinfo
  {volume} {782}},\ \bibinfo {pages} {238 } (\bibinfo {year}
  {2018})}\BibitemShut {NoStop}%
\bibitem [{\citenamefont {Deltuva\text{ \ }}\ and\ \citenamefont
  {Lazauskas}(2019{\natexlab{a}})}]{DeltuvaLazauskas2019PRC}%
  \BibitemOpen
  \bibfield  {author} {\bibinfo {author} {\bibfnamefont {A.}~\bibnamefont
  {Deltuva\text{ \ }}}and\ \bibinfo {author} {\bibfnamefont {R.}~\bibnamefont
  {Lazauskas}},\ }\bibfield  {title} {\bibinfo {title} {{Tetraneutron resonance
  in the presence of a dineutron}},\ }\href
  {https://doi.org/{10.1103/PhysRevC.100.044002}} {\bibfield  {journal}
  {\bibinfo  {journal} {{Phys. Rev. C}}\ }\textbf {\bibinfo {volume} {{100}}},\
  \bibinfo {pages} {044002} (\bibinfo {year}
  {{2019}}{\natexlab{a}})}\BibitemShut {NoStop}%
\bibitem [{\citenamefont {Deltuva\text{ \ }}\ and\ \citenamefont
  {Lazauskas}(2019{\natexlab{b}})}]{DeltuvaLazauskas2019commentPRL}%
  \BibitemOpen
  \bibfield  {author} {\bibinfo {author} {\bibfnamefont {A.}~\bibnamefont
  {Deltuva\text{ \ }}}and\ \bibinfo {author} {\bibfnamefont {R.}~\bibnamefont
  {Lazauskas}},\ }\bibfield  {title} {\bibinfo {title} {{Comment on ``Is a
  Trineutron Resonance Lower in Energy than a Tetraneutron Resonance?{''}}},\
  }\href {https://doi.org/{10.1103/PhysRevLett.123.069201}} {\bibfield
  {journal} {\bibinfo  {journal} {{Phys. Rev. Lett.}}\ }\textbf {\bibinfo
  {volume} {{123}}},\ \bibinfo {pages} {069201} (\bibinfo {year}
  {{2019}}{\natexlab{b}})}\BibitemShut {NoStop}%
\bibitem [{\citenamefont {Gandolfi}\ \emph {et~al.}(2019)\citenamefont
  {Gandolfi}, \citenamefont {Hammer}, \citenamefont {Klos}, \citenamefont
  {Lynn},\ and\ \citenamefont {Schwenk}}]{GandolfiHammer2019PRL}%
  \BibitemOpen
  \bibfield  {author} {\bibinfo {author} {\bibfnamefont {S.}~\bibnamefont
  {Gandolfi}}, \bibinfo {author} {\bibfnamefont {H.~W.}\ \bibnamefont
  {Hammer}}, \bibinfo {author} {\bibfnamefont {P.}~\bibnamefont {Klos}},
  \bibinfo {author} {\bibfnamefont {J.~E.}\ \bibnamefont {Lynn}}, and\ \bibinfo
  {author} {\bibfnamefont {A.}~\bibnamefont {Schwenk}},\ }\bibfield  {title}
  {\bibinfo {title} {{Comment on ``Is a Trineutron Resonance Lower in Energy
  than a Tetraneutron Resonance?{''} Reply}},\ }\href
  {https://doi.org/{10.1103/PhysRevLett.123.069202}} {\bibfield  {journal}
  {\bibinfo  {journal} {{Phys. Rev. Lett.}}\ }\textbf {\bibinfo {volume}
  {{123}}},\ \bibinfo {pages} {069202} (\bibinfo {year} {{2019}})}\BibitemShut
  {NoStop}%
\bibitem [{\citenamefont {Hiyama\text{ \ }}\ and\ \citenamefont
  {Kamimura}(2018)}]{HiyamaKamimura2018FPhys}%
  \BibitemOpen
  \bibfield  {author} {\bibinfo {author} {\bibfnamefont {E.}~\bibnamefont
  {Hiyama\text{ \ }}}and\ \bibinfo {author} {\bibfnamefont {M.}~\bibnamefont
  {Kamimura}},\ }\bibfield  {title} {\bibinfo {title} {{Study of various
  few-body systems using Gaussian expansion method (GEM)}},\ }\href
  {https://doi.org/{10.1007/s11467-018-0828-5}} {\bibfield  {journal} {\bibinfo
   {journal} {{Frontiers of Physics}}\ }\textbf {\bibinfo {volume} {{13}}},\
  \bibinfo {pages} {132106} (\bibinfo {year} {{2018}})}\BibitemShut {NoStop}%
\bibitem [{\citenamefont {{Botero, J. and Greene, C.
  H.}}(1986)}]{botero1986PRL}%
  \BibitemOpen
  \bibfield  {author} {\bibinfo {author} {\bibnamefont {{Botero, J. and Greene,
  C. H.}}},\ }\bibfield  {title} {{\selectlanguage {English}\bibinfo {title}
  {Resonant photodetachment of the positronium negative-ion}},\ }\href@noop {}
  {\bibfield  {journal} {\bibinfo  {journal} {Phys. Rev. Lett.}\ }\textbf
  {\bibinfo {volume} {56}},\ \bibinfo {pages} {1366} (\bibinfo {year}
  {1986})}\BibitemShut {NoStop}%
\bibitem [{\citenamefont {Lin}(1975)}]{lin1975PRL}%
  \BibitemOpen
  \bibfield  {author} {\bibinfo {author} {\bibfnamefont {C.~D.}\ \bibnamefont
  {Lin}},\ }\bibfield  {title} {{\selectlanguage {English}\bibinfo {title}
  {Feshbach and shape resonances in the e-{H} $^1\uppercase{P}^o$ system}},\
  }\href@noop {} {\bibfield  {journal} {\bibinfo  {journal} {Phys. Rev. Lett.}\
  }\textbf {\bibinfo {volume} {35}},\ \bibinfo {pages} {1150} (\bibinfo {year}
  {1975})}\BibitemShut {NoStop}%
\bibitem [{\citenamefont {Michishio}\ \emph {et~al.}(2016)\citenamefont
  {Michishio}, \citenamefont {Kanai}, \citenamefont {Kuma}, \citenamefont
  {Azuma}, \citenamefont {Wada}, \citenamefont {Mochizuki}, \citenamefont
  {Hyodo}, \citenamefont {Yagishita},\ and\ \citenamefont
  {Nagashima}}]{MichishioNagashima2016NatureComm}%
  \BibitemOpen
  \bibfield  {author} {\bibinfo {author} {\bibfnamefont {K.}~\bibnamefont
  {Michishio}}, \bibinfo {author} {\bibfnamefont {T.}~\bibnamefont {Kanai}},
  \bibinfo {author} {\bibfnamefont {S.}~\bibnamefont {Kuma}}, \bibinfo {author}
  {\bibfnamefont {T.}~\bibnamefont {Azuma}}, \bibinfo {author} {\bibfnamefont
  {K.}~\bibnamefont {Wada}}, \bibinfo {author} {\bibfnamefont {I.}~\bibnamefont
  {Mochizuki}}, \bibinfo {author} {\bibfnamefont {T.}~\bibnamefont {Hyodo}},
  \bibinfo {author} {\bibfnamefont {A.}~\bibnamefont {Yagishita}}, and\
  \bibinfo {author} {\bibfnamefont {Y.}~\bibnamefont {Nagashima}},\ }\bibfield
  {title} {\bibinfo {title} {{Observation of a shape resonance of the
  positronium negative ion}},\ }\href {https://doi.org/{10.1038/ncomms11060}}
  {\bibfield  {journal} {\bibinfo  {journal} {Nature Communications}\ }\textbf
  {\bibinfo {volume} {{7}}},\ \bibinfo {pages} {{11060}} (\bibinfo {year}
  {{2016}})}\BibitemShut {NoStop}%
\bibitem [{\citenamefont {Bryant}\ \emph {et~al.}(1977)\citenamefont {Bryant},
  \citenamefont {Dieterle}, \citenamefont {Donahue}, \citenamefont {Sharifian},
  \citenamefont {Tootoonchi}, \citenamefont {Wolfe}, \citenamefont {Gram},\
  and\ \citenamefont {Yates-Williams}}]{bryant1977PRL}%
  \BibitemOpen
  \bibfield  {author} {\bibinfo {author} {\bibfnamefont {H.~C.}\ \bibnamefont
  {Bryant}}, \bibinfo {author} {\bibfnamefont {B.~D.}\ \bibnamefont
  {Dieterle}}, \bibinfo {author} {\bibfnamefont {J.}~\bibnamefont {Donahue}},
  \bibinfo {author} {\bibfnamefont {H.}~\bibnamefont {Sharifian}}, \bibinfo
  {author} {\bibfnamefont {H.}~\bibnamefont {Tootoonchi}}, \bibinfo {author}
  {\bibfnamefont {D.~M.}\ \bibnamefont {Wolfe}}, \bibinfo {author}
  {\bibfnamefont {P.~A.~M.}\ \bibnamefont {Gram}}, and\ \bibinfo {author}
  {\bibfnamefont {M.~A.}\ \bibnamefont {Yates-Williams}},\ }\bibfield  {title}
  {\bibinfo {title} {{Observation of Resonances near 11 eV in the
  Photodetachment Cross Section of the H$^-$ Ion}},\ }\href@noop {} {\bibfield
  {journal} {\bibinfo  {journal} {Phys. Rev. Lett.}\ }\textbf {\bibinfo
  {volume} {38}},\ \bibinfo {pages} {228} (\bibinfo {year} {1977})}\BibitemShut
  {NoStop}%
\bibitem [{\citenamefont {Suzuki\text{ \ }}\ and\ \citenamefont
  {Varga}(1988)}]{SV2}%
  \BibitemOpen
  \bibfield  {author} {\bibinfo {author} {\bibfnamefont {Y.}~\bibnamefont
  {Suzuki\text{ \ }}}and\ \bibinfo {author} {\bibfnamefont {K.}~\bibnamefont
  {Varga}},\ }\href@noop {} {\emph {\bibinfo {title} {Stochastic Variational
  Approach to Quantum-Mechanical Few-Body Problems}}}\ (\bibinfo  {publisher}
  {Springer},\ \bibinfo {address} {Heidelberg},\ \bibinfo {year}
  {1988})\BibitemShut {NoStop}%
\bibitem [{\citenamefont {Rittenhouse}\ \emph {et~al.}(2011)\citenamefont
  {Rittenhouse}, \citenamefont {von Stecher}, \citenamefont {D'Incao},
  \citenamefont {Mehta},\ and\ \citenamefont {Greene}}]{Rittenhouse_2011}%
  \BibitemOpen
  \bibfield  {author} {\bibinfo {author} {\bibfnamefont {S.~T.}\ \bibnamefont
  {Rittenhouse}}, \bibinfo {author} {\bibfnamefont {J.}~\bibnamefont {von
  Stecher}}, \bibinfo {author} {\bibfnamefont {J.~P.}\ \bibnamefont {D'Incao}},
  \bibinfo {author} {\bibfnamefont {N.~P.}\ \bibnamefont {Mehta}}, and\
  \bibinfo {author} {\bibfnamefont {C.~H.}\ \bibnamefont {Greene}},\ }\bibfield
   {title} {\bibinfo {title} {The hyperspherical four-fermion problem},\ }\href
  {https://doi.org/10.1088/0953-4075/44/17/172001} {\bibfield  {journal}
  {\bibinfo  {journal} {Journal of Physics B: Atomic, Molecular and Optical
  Physics}\ }\textbf {\bibinfo {volume} {44}},\ \bibinfo {pages} {172001}
  (\bibinfo {year} {2011})}\BibitemShut {NoStop}%
\bibitem [{\citenamefont {Mitroy}\ \emph {et~al.}(2013)\citenamefont {Mitroy},
  \citenamefont {Bubin}, \citenamefont {Horiuchi}, \citenamefont {Suzuki},
  \citenamefont {Adamowicz}, \citenamefont {Cencek}, \citenamefont {Szalewicz},
  \citenamefont {Komasa}, \citenamefont {Blume},\ and\ \citenamefont
  {Varga}}]{RevModPhys.85.693}%
  \BibitemOpen
  \bibfield  {author} {\bibinfo {author} {\bibfnamefont {J.}~\bibnamefont
  {Mitroy}}, \bibinfo {author} {\bibfnamefont {S.}~\bibnamefont {Bubin}},
  \bibinfo {author} {\bibfnamefont {W.}~\bibnamefont {Horiuchi}}, \bibinfo
  {author} {\bibfnamefont {Y.}~\bibnamefont {Suzuki}}, \bibinfo {author}
  {\bibfnamefont {L.}~\bibnamefont {Adamowicz}}, \bibinfo {author}
  {\bibfnamefont {W.}~\bibnamefont {Cencek}}, \bibinfo {author} {\bibfnamefont
  {K.}~\bibnamefont {Szalewicz}}, \bibinfo {author} {\bibfnamefont
  {J.}~\bibnamefont {Komasa}}, \bibinfo {author} {\bibfnamefont
  {D.}~\bibnamefont {Blume}}, and\ \bibinfo {author} {\bibfnamefont
  {K.}~\bibnamefont {Varga}},\ }\bibfield  {title} {\bibinfo {title} {Theory
  and application of explicitly correlated gaussians},\ }\href
  {https://doi.org/10.1103/RevModPhys.85.693} {\bibfield  {journal} {\bibinfo
  {journal} {Rev. Mod. Phys.}\ }\textbf {\bibinfo {volume} {85}},\ \bibinfo
  {pages} {693} (\bibinfo {year} {2013})}\BibitemShut {NoStop}%
\bibitem [{\citenamefont {Varga}\ \emph {et~al.}(1998)\citenamefont {Varga},
  \citenamefont {Suzuki},\ and\ \citenamefont {Usukura}}]{Varga1998}%
  \BibitemOpen
  \bibfield  {author} {\bibinfo {author} {\bibfnamefont {K.}~\bibnamefont
  {Varga}}, \bibinfo {author} {\bibfnamefont {Y.}~\bibnamefont {Suzuki}}, and\
  \bibinfo {author} {\bibfnamefont {J.}~\bibnamefont {Usukura}},\ }\bibfield
  {title} {\bibinfo {title} {Global-vector representation of the angular motion
  of few-particle systems},\ }\href {https://doi.org/10.1007/s006010050077}
  {\bibfield  {journal} {\bibinfo  {journal} {Few-Body Systems}\ }\textbf
  {\bibinfo {volume} {24}},\ \bibinfo {pages} {81} (\bibinfo {year}
  {1998})}\BibitemShut {NoStop}%
\bibitem [{\citenamefont {Wang}(2012)}]{Wang}%
  \BibitemOpen
  \bibfield  {author} {\bibinfo {author} {\bibfnamefont {J.}~\bibnamefont
  {Wang}},\ }\emph {\bibinfo {title} {Hyperspherical Approach to Quantal
  Three-body Theory}},\ \href@noop {} {Ph.D. thesis},\ \bibinfo  {school}
  {University of Colorado, Boulder} (\bibinfo {year} {2012})\BibitemShut
  {NoStop}%
\bibitem [{\citenamefont {Daily}\ \emph {et~al.}(2015)\citenamefont {Daily},
  \citenamefont {von Stecher},\ and\ \citenamefont {Greene}}]{e_+e_-Daily}%
  \BibitemOpen
  \bibfield  {author} {\bibinfo {author} {\bibfnamefont {K.~M.}\ \bibnamefont
  {Daily}}, \bibinfo {author} {\bibfnamefont {J.}~\bibnamefont {von Stecher}},
  and\ \bibinfo {author} {\bibfnamefont {C.~H.}\ \bibnamefont {Greene}},\
  }\bibfield  {title} {\bibinfo {title} {Scattering properties of the
  $2{e}^{\ensuremath{-}}2{e}^{+}$ polyelectronic system},\ }\href
  {https://doi.org/10.1103/PhysRevA.91.012512} {\bibfield  {journal} {\bibinfo
  {journal} {Phys. Rev. A}\ }\textbf {\bibinfo {volume} {91}},\ \bibinfo
  {pages} {012512} (\bibinfo {year} {2015})}\BibitemShut {NoStop}%
\bibitem [{\citenamefont {Daily}(2015)}]{DailyAsym}%
  \BibitemOpen
  \bibfield  {author} {\bibinfo {author} {\bibfnamefont {K.~M.}\ \bibnamefont
  {Daily}},\ }\bibfield  {title} {\bibinfo {title} {Hyperspherical asymptotics
  of a system of four charged particles},\ }\href
  {https://doi.org/10.1007/s00601-015-0979-7} {\bibfield  {journal} {\bibinfo
  {journal} {Few-Body Systems}\ }\textbf {\bibinfo {volume} {56}},\ \bibinfo
  {pages} {809} (\bibinfo {year} {2015})}\BibitemShut {NoStop}%
\bibitem [{\citenamefont {Wang}\ \emph {et~al.}(2012)\citenamefont {Wang},
  \citenamefont {D'Incao}, \citenamefont {Wang},\ and\ \citenamefont
  {Greene}}]{PhysRevA.86.062511}%
  \BibitemOpen
  \bibfield  {author} {\bibinfo {author} {\bibfnamefont {J.}~\bibnamefont
  {Wang}}, \bibinfo {author} {\bibfnamefont {J.~P.}\ \bibnamefont {D'Incao}},
  \bibinfo {author} {\bibfnamefont {Y.}~\bibnamefont {Wang}}, and\ \bibinfo
  {author} {\bibfnamefont {C.~H.}\ \bibnamefont {Greene}},\ }\bibfield  {title}
  {\bibinfo {title} {Universal three-body recombination via resonant $d$-wave
  interactions},\ }\href {https://doi.org/10.1103/PhysRevA.86.062511}
  {\bibfield  {journal} {\bibinfo  {journal} {Phys. Rev. A}\ }\textbf {\bibinfo
  {volume} {86}},\ \bibinfo {pages} {062511} (\bibinfo {year}
  {2012})}\BibitemShut {NoStop}%
\bibitem [{\citenamefont {Wiringa}\ \emph {et~al.}(1995)\citenamefont
  {Wiringa}, \citenamefont {Stoks},\ and\ \citenamefont
  {Schiavilla}}]{Wiringa:1994w}%
  \BibitemOpen
  \bibfield  {author} {\bibinfo {author} {\bibfnamefont {R.~B.}\ \bibnamefont
  {Wiringa}}, \bibinfo {author} {\bibfnamefont {V.}~\bibnamefont {Stoks}}, and\
  \bibinfo {author} {\bibfnamefont {R.}~\bibnamefont {Schiavilla}},\ }\bibfield
   {title} {\bibinfo {title} {{An Accurate nucleon-nucleon potential with
  charge independence breaking}},\ }\href
  {https://doi.org/10.1103/PhysRevC.51.38} {\bibfield  {journal} {\bibinfo
  {journal} {Phys. Rev. C}\ }\textbf {\bibinfo {volume} {51}},\ \bibinfo
  {pages} {38} (\bibinfo {year} {1995})},\ \Eprint
  {https://arxiv.org/abs/nucl-th/9408016} {arXiv:nucl-th/9408016} \BibitemShut
  {NoStop}%
\bibitem [{\citenamefont {Piarulli}\ \emph {et~al.}(2016)\citenamefont
  {Piarulli}, \citenamefont {Girlanda}, \citenamefont {Schiavilla},
  \citenamefont {Kievsky}, \citenamefont {Lovato}, \citenamefont {Marcucci},
  \citenamefont {Pieper}, \citenamefont {Viviani},\ and\ \citenamefont
  {Wiringa}}]{Piarulli:2016vel}%
  \BibitemOpen
  \bibfield  {author} {\bibinfo {author} {\bibfnamefont {M.}~\bibnamefont
  {Piarulli}}, \bibinfo {author} {\bibfnamefont {L.}~\bibnamefont {Girlanda}},
  \bibinfo {author} {\bibfnamefont {R.}~\bibnamefont {Schiavilla}}, \bibinfo
  {author} {\bibfnamefont {A.}~\bibnamefont {Kievsky}}, \bibinfo {author}
  {\bibfnamefont {A.}~\bibnamefont {Lovato}}, \bibinfo {author} {\bibfnamefont
  {L.~E.}\ \bibnamefont {Marcucci}}, \bibinfo {author} {\bibfnamefont {S.~C.}\
  \bibnamefont {Pieper}}, \bibinfo {author} {\bibfnamefont {M.}~\bibnamefont
  {Viviani}}, and\ \bibinfo {author} {\bibfnamefont {R.~B.}\ \bibnamefont
  {Wiringa}},\ }\bibfield  {title} {\bibinfo {title} {{Local chiral potentials
  with $\Delta$-intermediate states and the structure of light nuclei}},\
  }\href {https://doi.org/10.1103/PhysRevC.94.054007} {\bibfield  {journal}
  {\bibinfo  {journal} {Phys. Rev. C}\ }\textbf {\bibinfo {volume} {94}},\
  \bibinfo {pages} {054007} (\bibinfo {year} {2016})},\ \Eprint
  {https://arxiv.org/abs/1606.06335} {arXiv:1606.06335 [nucl-th]} \BibitemShut
  {NoStop}%
\bibitem [{\citenamefont {Baroni}\ \emph {et~al.}(2018)\citenamefont {Baroni}
  \emph {et~al.}}]{Baroni:2018fdn}%
  \BibitemOpen
  \bibfield  {author} {\bibinfo {author} {\bibfnamefont {A.}~\bibnamefont
  {Baroni}} \emph {et~al.},\ }\bibfield  {title} {\bibinfo {title} {{Local
  chiral interactions, the tritium Gamow-Teller matrix element, and the
  three-nucleon contact term}},\ }\href
  {https://doi.org/10.1103/PhysRevC.98.044003} {\bibfield  {journal} {\bibinfo
  {journal} {Phys. Rev. C}\ }\textbf {\bibinfo {volume} {98}},\ \bibinfo
  {pages} {044003} (\bibinfo {year} {2018})},\ \Eprint
  {https://arxiv.org/abs/1806.10245} {arXiv:1806.10245 [nucl-th]} \BibitemShut
  {NoStop}%
\bibitem [{\citenamefont {D'Incao}(2018)}]{DIncaoReview}%
  \BibitemOpen
  \bibfield  {author} {\bibinfo {author} {\bibfnamefont {J.~P.}\ \bibnamefont
  {D'Incao}},\ }\bibfield  {title} {\bibinfo {title} {{Few-body physics in
  resonantly interacting ultracold quantum gases}},\ }\href
  {https://doi.org/{10.1088/1361-6455/aaa116}} {\bibfield  {journal} {\bibinfo
  {journal} {{J. Phys. B}}\ }\textbf {\bibinfo {volume} {{51}}},\ \bibinfo
  {pages} {{043001}} (\bibinfo {year} {{2018}})}\BibitemShut {NoStop}%
\bibitem [{\citenamefont {Greene}\ \emph {et~al.}(2017)\citenamefont {Greene},
  \citenamefont {Giannakeas},\ and\ \citenamefont
  {P\'erez-R\'{\i}os}}]{Greene2017RMP}%
  \BibitemOpen
  \bibfield  {author} {\bibinfo {author} {\bibfnamefont {C.~H.}\ \bibnamefont
  {Greene}}, \bibinfo {author} {\bibfnamefont {P.}~\bibnamefont {Giannakeas}},
  and\ \bibinfo {author} {\bibfnamefont {J.}~\bibnamefont
  {P\'erez-R\'{\i}os}},\ }\bibfield  {title} {\bibinfo {title} {Universal
  few-body physics and cluster formation},\ }\href
  {https://doi.org/10.1103/RevModPhys.89.035006} {\bibfield  {journal}
  {\bibinfo  {journal} {Rev. Mod. Phys.}\ }\textbf {\bibinfo {volume} {89}},\
  \bibinfo {pages} {035006} (\bibinfo {year} {2017})}\BibitemShut {NoStop}%
\bibitem [{\citenamefont {Naidon\text{ \ }}\ and\ \citenamefont
  {Endo}(2017)}]{NaidonReview}%
  \BibitemOpen
  \bibfield  {author} {\bibinfo {author} {\bibfnamefont {P.}~\bibnamefont
  {Naidon\text{ \ }}}and\ \bibinfo {author} {\bibfnamefont {S.}~\bibnamefont
  {Endo}},\ }\bibfield  {title} {\bibinfo {title} {{Efimov physics: a
  review}},\ }\href {https://doi.org/{10.1088/1361-6633/aa50e8}} {\bibfield
  {journal} {\bibinfo  {journal} {{Reports on Progress in Physics}}\ }\textbf
  {\bibinfo {volume} {{80}}},\ \bibinfo {pages} {056001} (\bibinfo {year}
  {{2017}})}\BibitemShut {NoStop}%
\bibitem [{\citenamefont {Suno\text{ \ }}\ and\ \citenamefont
  {Esry}(2008)}]{PhysRevA.78.062701}%
  \BibitemOpen
  \bibfield  {author} {\bibinfo {author} {\bibfnamefont {H.}~\bibnamefont
  {Suno\text{ \ }}}and\ \bibinfo {author} {\bibfnamefont {B.~D.}\ \bibnamefont
  {Esry}},\ }\bibfield  {title} {\bibinfo {title} {Adiabatic hyperspherical
  study of triatomic helium systems},\ }\href
  {https://doi.org/10.1103/PhysRevA.78.062701} {\bibfield  {journal} {\bibinfo
  {journal} {Phys. Rev. A}\ }\textbf {\bibinfo {volume} {78}},\ \bibinfo
  {pages} {062701} (\bibinfo {year} {2008})}\BibitemShut {NoStop}%
\bibitem [{\citenamefont {Hagen}\ \emph {et~al.}(2013)\citenamefont {Hagen},
  \citenamefont {Hagen}, \citenamefont {Hammer},\ and\ \citenamefont
  {Platter}}]{PhysRevLett.111.132501}%
  \BibitemOpen
  \bibfield  {author} {\bibinfo {author} {\bibfnamefont {G.}~\bibnamefont
  {Hagen}}, \bibinfo {author} {\bibfnamefont {P.}~\bibnamefont {Hagen}},
  \bibinfo {author} {\bibfnamefont {H.-W.}\ \bibnamefont {Hammer}}, and\
  \bibinfo {author} {\bibfnamefont {L.}~\bibnamefont {Platter}},\ }\bibfield
  {title} {\bibinfo {title} {Efimov physics around the neutron-rich
  $^{60}\mathrm{Ca}$ isotope},\ }\href
  {https://doi.org/10.1103/PhysRevLett.111.132501} {\bibfield  {journal}
  {\bibinfo  {journal} {Phys. Rev. Lett.}\ }\textbf {\bibinfo {volume} {111}},\
  \bibinfo {pages} {132501} (\bibinfo {year} {2013})}\BibitemShut {NoStop}%
\bibitem [{\citenamefont {Efimov\text{ \ }}\ and\ \citenamefont
  {Tkachenko}(1988)}]{Efimov1988}%
  \BibitemOpen
  \bibfield  {author} {\bibinfo {author} {\bibfnamefont {V.}~\bibnamefont
  {Efimov\text{ \ }}}and\ \bibinfo {author} {\bibfnamefont {E.~G.}\
  \bibnamefont {Tkachenko}},\ }\bibfield  {title} {\bibinfo {title} {On the
  correlation between the triton binding energy and the neutron-deuteron
  doublet scattering length},\ }\href {https://doi.org/10.1007/BF01076330}
  {\bibfield  {journal} {\bibinfo  {journal} {Few-Body Systems}\ }\textbf
  {\bibinfo {volume} {4}},\ \bibinfo {pages} {71} (\bibinfo {year}
  {1988})}\BibitemShut {NoStop}%
\bibitem [{\citenamefont {Kievsky\text{ \ }}\ and\ \citenamefont
  {Gattobigio}(2016)}]{KievskyandG2016}%
  \BibitemOpen
  \bibfield  {author} {\bibinfo {author} {\bibfnamefont {A.}~\bibnamefont
  {Kievsky\text{ \ }}}and\ \bibinfo {author} {\bibfnamefont {M.}~\bibnamefont
  {Gattobigio}},\ }\bibfield  {title} {\bibinfo {title} {Efimov physics with
  ${1/2}$ spin-isospin fermions},\ }\href
  {https://doi.org/10.1007/s00601-016-1049-5} {\bibfield  {journal} {\bibinfo
  {journal} {Few-Body Systems}\ }\textbf {\bibinfo {volume} {57}},\ \bibinfo
  {pages} {217} (\bibinfo {year} {2016})}\BibitemShut {NoStop}%
\bibitem [{\citenamefont {Gattobigio}\ \emph {et~al.}(2019)\citenamefont
  {Gattobigio}, \citenamefont {Kievsky},\ and\ \citenamefont
  {Viviani}}]{PhysRevC.100.034004}%
  \BibitemOpen
  \bibfield  {author} {\bibinfo {author} {\bibfnamefont {M.}~\bibnamefont
  {Gattobigio}}, \bibinfo {author} {\bibfnamefont {A.}~\bibnamefont {Kievsky}},
  and\ \bibinfo {author} {\bibfnamefont {M.}~\bibnamefont {Viviani}},\
  }\bibfield  {title} {\bibinfo {title} {Embedding nuclear physics inside the
  unitary-limit window},\ }\href {https://doi.org/10.1103/PhysRevC.100.034004}
  {\bibfield  {journal} {\bibinfo  {journal} {Phys. Rev. C}\ }\textbf {\bibinfo
  {volume} {100}},\ \bibinfo {pages} {034004} (\bibinfo {year}
  {2019})}\BibitemShut {NoStop}%
\bibitem [{\citenamefont {Kievsky}\ \emph {et~al.}(2017)\citenamefont
  {Kievsky}, \citenamefont {Viviani}, \citenamefont {Gattobigio},\ and\
  \citenamefont {Girlanda}}]{PhysRevC.95.024001}%
  \BibitemOpen
  \bibfield  {author} {\bibinfo {author} {\bibfnamefont {A.}~\bibnamefont
  {Kievsky}}, \bibinfo {author} {\bibfnamefont {M.}~\bibnamefont {Viviani}},
  \bibinfo {author} {\bibfnamefont {M.}~\bibnamefont {Gattobigio}}, and\
  \bibinfo {author} {\bibfnamefont {L.}~\bibnamefont {Girlanda}},\ }\bibfield
  {title} {\bibinfo {title} {Implications of efimov physics for the description
  of three and four nucleons in chiral effective field theory},\ }\href
  {https://doi.org/10.1103/PhysRevC.95.024001} {\bibfield  {journal} {\bibinfo
  {journal} {Phys. Rev. C}\ }\textbf {\bibinfo {volume} {95}},\ \bibinfo
  {pages} {024001} (\bibinfo {year} {2017})}\BibitemShut {NoStop}%
\bibitem [{\citenamefont {Kievsky}\ \emph {et~al.}(2018)\citenamefont
  {Kievsky}, \citenamefont {Viviani}, \citenamefont {Logoteta}, \citenamefont
  {Bombaci},\ and\ \citenamefont {Girlanda}}]{PhysRevLett.121.072701}%
  \BibitemOpen
  \bibfield  {author} {\bibinfo {author} {\bibfnamefont {A.}~\bibnamefont
  {Kievsky}}, \bibinfo {author} {\bibfnamefont {M.}~\bibnamefont {Viviani}},
  \bibinfo {author} {\bibfnamefont {D.}~\bibnamefont {Logoteta}}, \bibinfo
  {author} {\bibfnamefont {I.}~\bibnamefont {Bombaci}}, and\ \bibinfo {author}
  {\bibfnamefont {L.}~\bibnamefont {Girlanda}},\ }\bibfield  {title} {\bibinfo
  {title} {Correlations imposed by the unitary limit between few-nucleon
  systems, nuclear matter, and neutron stars},\ }\href
  {https://doi.org/10.1103/PhysRevLett.121.072701} {\bibfield  {journal}
  {\bibinfo  {journal} {Phys. Rev. Lett.}\ }\textbf {\bibinfo {volume} {121}},\
  \bibinfo {pages} {072701} (\bibinfo {year} {2018})}\BibitemShut {NoStop}%
\bibitem [{\citenamefont {Regal\text{ \ }}\ and\ \citenamefont
  {Jin}(2007)}]{regal2007}%
  \BibitemOpen
  \bibfield  {author} {\bibinfo {author} {\bibfnamefont {C.~A.}\ \bibnamefont
  {Regal\text{ \ }}}and\ \bibinfo {author} {\bibfnamefont {D.~S.}\ \bibnamefont
  {Jin}},\ }\bibfield  {title} {\bibinfo {title} {Experimental realization of
  the {BCS}-{BEC} crossover with a {F}ermi gas of atoms},\ }\href@noop {}
  {\bibfield  {journal} {\bibinfo  {journal} {Adv. At. Mol. Opt. Phys.}\
  }\textbf {\bibinfo {volume} {54}},\ \bibinfo {pages} {1} (\bibinfo {year}
  {2007})}\BibitemShut {NoStop}%
\bibitem [{\citenamefont {von Stecher\text{ \ }}\ and\ \citenamefont
  {Greene}(2007)}]{PhysRevLett.99.090402}%
  \BibitemOpen
  \bibfield  {author} {\bibinfo {author} {\bibfnamefont {J.}~\bibnamefont {von
  Stecher\text{ \ }}}and\ \bibinfo {author} {\bibfnamefont {C.~H.}\
  \bibnamefont {Greene}},\ }\bibfield  {title} {\bibinfo {title} {Spectrum and
  dynamics of the \uppercase{BCS-BEC} crossover from a few-body perspective},\
  }\href {https://doi.org/10.1103/PhysRevLett.99.090402} {\bibfield  {journal}
  {\bibinfo  {journal} {Phys. Rev. Lett.}\ }\textbf {\bibinfo {volume} {99}},\
  \bibinfo {pages} {090402} (\bibinfo {year} {2007})}\BibitemShut {NoStop}%
\bibitem [{\citenamefont {Pudliner}\ \emph {et~al.}(1997)\citenamefont
  {Pudliner}, \citenamefont {Pandharipande}, \citenamefont {Carlson},
  \citenamefont {Pieper},\ and\ \citenamefont {Wiringa}}]{UIX}%
  \BibitemOpen
  \bibfield  {author} {\bibinfo {author} {\bibfnamefont {B.~S.}\ \bibnamefont
  {Pudliner}}, \bibinfo {author} {\bibfnamefont {V.~R.}\ \bibnamefont
  {Pandharipande}}, \bibinfo {author} {\bibfnamefont {J.}~\bibnamefont
  {Carlson}}, \bibinfo {author} {\bibfnamefont {S.~C.}\ \bibnamefont {Pieper}},
  and\ \bibinfo {author} {\bibfnamefont {R.~B.}\ \bibnamefont {Wiringa}},\
  }\bibfield  {title} {\bibinfo {title} {{Quantum Monte Carlo calculations of
  nuclei with $A<7$}},\ }\href {https://doi.org/10.1103/PhysRevC.56.1720}
  {\bibfield  {journal} {\bibinfo  {journal} {Phys. Rev. C}\ }\textbf {\bibinfo
  {volume} {56}},\ \bibinfo {pages} {1720} (\bibinfo {year}
  {1997})}\BibitemShut {NoStop}%
\bibitem [{\citenamefont {Pieper}\ \emph {et~al.}(2001)\citenamefont {Pieper},
  \citenamefont {Pandharipande}, \citenamefont {Wiringa},\ and\ \citenamefont
  {Carlson}}]{UIX2}%
  \BibitemOpen
  \bibfield  {author} {\bibinfo {author} {\bibfnamefont {S.}~\bibnamefont
  {Pieper}}, \bibinfo {author} {\bibfnamefont {V.}~\bibnamefont
  {Pandharipande}}, \bibinfo {author} {\bibfnamefont {R.}~\bibnamefont
  {Wiringa}}, and\ \bibinfo {author} {\bibfnamefont {J.}~\bibnamefont
  {Carlson}},\ }\bibfield  {title} {\bibinfo {title} {{Realistic models of
  pion-exchange three-nucleon interactions}},\ }\href
  {https://doi.org/{10.1103/PhysRevC.64.014001}} {\bibfield  {journal}
  {\bibinfo  {journal} {{Phys. Rev. C}}\ }\textbf {\bibinfo {volume} {{64}}},\
  \bibinfo {pages} {014001} (\bibinfo {year} {{2001}})}\BibitemShut {NoStop}%
\bibitem [{\citenamefont {Carlson}\ \emph {et~al.}(1983)\citenamefont
  {Carlson}, \citenamefont {Pandharipande},\ and\ \citenamefont
  {Wiringa}}]{Illinois3BF}%
  \BibitemOpen
  \bibfield  {author} {\bibinfo {author} {\bibfnamefont {J.}~\bibnamefont
  {Carlson}}, \bibinfo {author} {\bibfnamefont {V.}~\bibnamefont
  {Pandharipande}}, and\ \bibinfo {author} {\bibfnamefont {R.}~\bibnamefont
  {Wiringa}},\ }\bibfield  {title} {\bibinfo {title} {{3-Nucleon Interaction in
  3-Body, 4-Body and Infinity-Body Systems}},\ }\href@noop {} {\bibfield
  {journal} {\bibinfo  {journal} {{Nuclear Physics A}}\ }\textbf {\bibinfo
  {volume} {401}},\ \bibinfo {pages} {59} (\bibinfo {year} {1983})}\BibitemShut
  {NoStop}%
\bibitem [{\citenamefont {Pieper}(2008)}]{PieperIL7}%
  \BibitemOpen
  \bibfield  {author} {\bibinfo {author} {\bibfnamefont {S.~C.}\ \bibnamefont
  {Pieper}},\ }\bibfield  {title} {\bibinfo {title} {The \uppercase{I}llinois
  extension to the \uppercase{F}ujita‐\uppercase{M}iyazawa three‐nucleon
  force},\ }\href {https://doi.org/10.1063/1.2932280} {\bibfield  {journal}
  {\bibinfo  {journal} {AIP Conference Proceedings}\ }\textbf {\bibinfo
  {volume} {1011}},\ \bibinfo {pages} {143} (\bibinfo {year} {2008})},\ \Eprint
  {https://arxiv.org/abs/https://aip.scitation.org/doi/pdf/10.1063/1.2932280}
  {https://aip.scitation.org/doi/pdf/10.1063/1.2932280} \BibitemShut {NoStop}%
\bibitem [{\citenamefont {Piarulli}\ \emph {et~al.}(2018)\citenamefont
  {Piarulli}, \citenamefont {Baroni}, \citenamefont {Girlanda}, \citenamefont
  {Kievsky}, \citenamefont {Lovato}, \citenamefont {Lusk}, \citenamefont
  {Marcucci}, \citenamefont {Pieper}, \citenamefont {Schiavilla}, \citenamefont
  {Viviani},\ and\ \citenamefont {Wiringa}}]{PhysRevLett.120.052503}%
  \BibitemOpen
  \bibfield  {author} {\bibinfo {author} {\bibfnamefont {M.}~\bibnamefont
  {Piarulli}}, \bibinfo {author} {\bibfnamefont {A.}~\bibnamefont {Baroni}},
  \bibinfo {author} {\bibfnamefont {L.}~\bibnamefont {Girlanda}}, \bibinfo
  {author} {\bibfnamefont {A.}~\bibnamefont {Kievsky}}, \bibinfo {author}
  {\bibfnamefont {A.}~\bibnamefont {Lovato}}, \bibinfo {author} {\bibfnamefont
  {E.}~\bibnamefont {Lusk}}, \bibinfo {author} {\bibfnamefont {L.~E.}\
  \bibnamefont {Marcucci}}, \bibinfo {author} {\bibfnamefont {S.~C.}\
  \bibnamefont {Pieper}}, \bibinfo {author} {\bibfnamefont {R.}~\bibnamefont
  {Schiavilla}}, \bibinfo {author} {\bibfnamefont {M.}~\bibnamefont {Viviani}},
  and\ \bibinfo {author} {\bibfnamefont {R.~B.}\ \bibnamefont {Wiringa}},\
  }\bibfield  {title} {\bibinfo {title} {Light-nuclei spectra from chiral
  dynamics},\ }\href {https://doi.org/10.1103/PhysRevLett.120.052503}
  {\bibfield  {journal} {\bibinfo  {journal} {Phys. Rev. Lett.}\ }\textbf
  {\bibinfo {volume} {120}},\ \bibinfo {pages} {052503} (\bibinfo {year}
  {2018})}\BibitemShut {NoStop}%
\bibitem [{\citenamefont {Zhen\text{ \ }}\ and\ \citenamefont
  {Macek}(1988)}]{PhysRevA.38.1193}%
  \BibitemOpen
  \bibfield  {author} {\bibinfo {author} {\bibfnamefont {Z.}~\bibnamefont
  {Zhen\text{ \ }}}and\ \bibinfo {author} {\bibfnamefont {J.}~\bibnamefont
  {Macek}},\ }\bibfield  {title} {\bibinfo {title} {Loosely bound states of
  three particles},\ }\href {https://doi.org/10.1103/PhysRevA.38.1193}
  {\bibfield  {journal} {\bibinfo  {journal} {Phys. Rev. A}\ }\textbf {\bibinfo
  {volume} {38}},\ \bibinfo {pages} {1193} (\bibinfo {year}
  {1988})}\BibitemShut {NoStop}%
\bibitem [{\citenamefont {Esry\text{ \ }}\ and\ \citenamefont
  {Greene}(1999)}]{PhysRevA.60.1451}%
  \BibitemOpen
  \bibfield  {author} {\bibinfo {author} {\bibfnamefont {B.~D.}\ \bibnamefont
  {Esry\text{ \ }}}and\ \bibinfo {author} {\bibfnamefont {C.~H.}\ \bibnamefont
  {Greene}},\ }\bibfield  {title} {\bibinfo {title} {Validity of the
  shape-independent approximation for \uppercase{B}ose-\uppercase{E}instein
  condensates},\ }\href {https://doi.org/10.1103/PhysRevA.60.1451} {\bibfield
  {journal} {\bibinfo  {journal} {Phys. Rev. A}\ }\textbf {\bibinfo {volume}
  {60}},\ \bibinfo {pages} {1451} (\bibinfo {year} {1999})}\BibitemShut
  {NoStop}%
\bibitem [{\citenamefont {Yin\text{ \ }}\ and\ \citenamefont
  {Blume}(2015)}]{YinBlume2015pra}%
  \BibitemOpen
  \bibfield  {author} {\bibinfo {author} {\bibfnamefont {X.~Y.}\ \bibnamefont
  {Yin\text{ \ }}}and\ \bibinfo {author} {\bibfnamefont {D.}~\bibnamefont
  {Blume}},\ }\bibfield  {title} {\bibinfo {title} {{Trapped unitary
  two-component Fermi gases with up to ten particles}},\ }\href
  {https://doi.org/{10.1103/PhysRevA.92.013608}} {\bibfield  {journal}
  {\bibinfo  {journal} {{Phys. Rev. A}}\ }\textbf {\bibinfo {volume} {{92}}},\
  \bibinfo {pages} {013608} (\bibinfo {year} {{2015}})}\BibitemShut {NoStop}%
\bibitem [{\citenamefont {Blume}\ \emph {et~al.}(2007)\citenamefont {Blume},
  \citenamefont {Von~Stecher},\ and\ \citenamefont
  {Greene}}]{vonStecher2007prl}%
  \BibitemOpen
  \bibfield  {author} {\bibinfo {author} {\bibfnamefont {D.}~\bibnamefont
  {Blume}}, \bibinfo {author} {\bibfnamefont {J.}~\bibnamefont {Von~Stecher}},
  and\ \bibinfo {author} {\bibfnamefont {C.~H.}\ \bibnamefont {Greene}},\
  }\bibfield  {title} {\bibinfo {title} {{Universal properties of a trapped
  two-component Fermi gas at unitarity}},\ }\href
  {https://doi.org/{10.1103/PhysRevLett.99.233201}} {\bibfield  {journal}
  {\bibinfo  {journal} {{Phys. Rev. Lett.}}\ }\textbf {\bibinfo {volume}
  {{99}}},\ \bibinfo {pages} {{233201}} (\bibinfo {year} {{2007}})}\BibitemShut
  {NoStop}%
\bibitem [{\citenamefont {Werner\text{ \ }}\ and\ \citenamefont
  {Castin}(2006)}]{PhysRevLett.97.150401}%
  \BibitemOpen
  \bibfield  {author} {\bibinfo {author} {\bibfnamefont {F.}~\bibnamefont
  {Werner\text{ \ }}}and\ \bibinfo {author} {\bibfnamefont {Y.}~\bibnamefont
  {Castin}},\ }\bibfield  {title} {\bibinfo {title} {Unitary quantum three-body
  problem in a harmonic trap},\ }\href
  {https://doi.org/10.1103/PhysRevLett.97.150401} {\bibfield  {journal}
  {\bibinfo  {journal} {Phys. Rev. Lett.}\ }\textbf {\bibinfo {volume} {97}},\
  \bibinfo {pages} {150401} (\bibinfo {year} {2006})}\BibitemShut {NoStop}%
\bibitem [{\citenamefont {von Stecher\text{ \ }}\ and\ \citenamefont
  {Greene}(2009)}]{stecher2009PRA}%
  \BibitemOpen
  \bibfield  {author} {\bibinfo {author} {\bibfnamefont {J.}~\bibnamefont {von
  Stecher\text{ \ }}}and\ \bibinfo {author} {\bibfnamefont {C.~H.}\
  \bibnamefont {Greene}},\ }\bibfield  {title} {{\selectlanguage
  {English}\bibinfo {title} {Correlated {G}aussian hyperspherical method for
  few-body systems}},\ }\href@noop {} {\bibfield  {journal} {\bibinfo
  {journal} {Phys. Rev. A}\ }\textbf {\bibinfo {volume} {80}},\ \bibinfo
  {pages} {022504} (\bibinfo {year} {2009})}\BibitemShut {NoStop}%
\bibitem [{\citenamefont {Smith}(1960)}]{Smith1960PR}%
  \BibitemOpen
  \bibfield  {author} {\bibinfo {author} {\bibfnamefont {F.~T.}\ \bibnamefont
  {Smith}},\ }\bibfield  {title} {\bibinfo {title} {Lifetime matrix in
  collision theory},\ }\href {https://doi.org/10.1103/PhysRev.118.349}
  {\bibfield  {journal} {\bibinfo  {journal} {Phys. Rev.}\ }\textbf {\bibinfo
  {volume} {118}},\ \bibinfo {pages} {349} (\bibinfo {year}
  {1960})}\BibitemShut {NoStop}%
\bibitem [{\citenamefont {Texier}(2016)}]{TEXIER201616}%
  \BibitemOpen
  \bibfield  {author} {\bibinfo {author} {\bibfnamefont {C.}~\bibnamefont
  {Texier}},\ }\bibfield  {title} {\bibinfo {title} {Wigner time delay and
  related concepts: Application to transport in coherent conductors},\ }\href
  {https://doi.org/https://doi.org/10.1016/j.physe.2015.09.041} {\bibfield
  {journal} {\bibinfo  {journal} {Physica E: Low-dimensional Systems and
  Nanostructures}\ }\textbf {\bibinfo {volume} {82}},\ \bibinfo {pages} {16 }
  (\bibinfo {year} {2016})},\ \bibinfo {note} {frontiers in quantum electronic
  transport - In memory of Markus Büttiker}\BibitemShut {NoStop}%
\bibitem [{\citenamefont {Aymar}\ \emph {et~al.}(1996)\citenamefont {Aymar},
  \citenamefont {Greene},\ and\ \citenamefont {Luc-Koenig}}]{OrangeReview}%
  \BibitemOpen
  \bibfield  {author} {\bibinfo {author} {\bibfnamefont {M.}~\bibnamefont
  {Aymar}}, \bibinfo {author} {\bibfnamefont {C.~H.}\ \bibnamefont {Greene}},
  and\ \bibinfo {author} {\bibfnamefont {E.}~\bibnamefont {Luc-Koenig}},\
  }\bibfield  {title} {\bibinfo {title} {Multichannel {R}ydberg spectroscopy of
  complex atoms},\ }\href@noop {} {\bibfield  {journal} {\bibinfo  {journal}
  {Rev. Mod. Phys.}\ }\textbf {\bibinfo {volume} {68}},\ \bibinfo {pages}
  {1015} (\bibinfo {year} {1996})}\BibitemShut {NoStop}%
\end{thebibliography}%

\end{document}